\documentclass[a4paper,fleqn,usenatbib]{mnras}

\usepackage[T1]{fontenc}
\usepackage{ae,aecompl}

\usepackage{float}

\usepackage{graphicx}	% Including figure files
\usepackage{amsmath}	% Advanced maths commands
\usepackage{amssymb}	% Extra maths symbols
\usepackage{lineno}
\usepackage{afterpage}

\usepackage{newtxtext,newtxmath}

\setcitestyle{notesep={}}
\title[S0 Formation Pathways]{The two formation pathways of S0 galaxies}

\author[S. Deeley et al.]{
Simon Deeley$^{1}$\thanks{E-mail: s.deeley@uq.edu.au},
Michael J. Drinkwater$^{1}$, Sarah M. Sweet$^{1,2,3}$, Kenji Bekki$^{4}$, Warrick J. Couch$^{2}$, \newauthor Duncan A. Forbes$^{2}$, Arianna Dolfi$^{2}$\\
${}^1$School of Mathematics and Physics, University of Queensland, Brisbane, Queensland 4072, Australia\\
${}^2$Centre for Astrophysics \& Supercomputing, Swinburne University, Hawthorn, VIC 3122, Australia \\
${}^3$ARC Centre of Excellence for All Sky Astrophysics in 3 Dimensions (ASTRO 3D)\\
${}^4$International Centre for Radio Astronomy Research, The University of Western Australia, 35 Stirling Highway, Crawley, Western Australia, 6009, Australia \\
}

\date{Accepted XXX. Received YYY; in original form ZZZ}

\pubyear{2019}

\begin{document}
%\linenumbers
\label{firstpage}
\pagerange{\pageref{firstpage}--\pageref{lastpage}}
\maketitle
% Abstract of the paper  NOTE 250 word limit ****
\begin{abstract}

Despite their ubiquity throughout the Universe, the formation of S0 galaxies remains uncertain. Recent observations have revealed that S0 galaxies make up a diverse population which is difficult to explain with a single formation pathway, suggesting that the picture of how these galaxies form is more complicated than originally envisioned. Here we take advantage of the latest hydrodynamical cosmological simulations and follow up these studies with an investigation into the formation histories of S0s in IllustrisTNG. We first classify IllustrisTNG galaxies in a way which is fully consistent with the observations, and reproduce the observed photometric and environmental distributions seen for the S0 population. We then trace the formation histories of S0 galaxies back through time, identifying two main distinct pathways; those which experienced gas stripping via group infalls (37 percent of S0s) or significant merger events (57 percent). We find that those forming via mergers feature a transient star-forming ring, whose present-day occurrence rate matches observations. We find that these formation pathways together can reproduce the range in rotational support in observed S0s, concluding that there are two main formation pathways for S0 galaxies. 

\end{abstract}


%\begin{keywords}

%\end{keywords}

%%%%%%%%%%%%%%%%%%%%%%%%%%%%%%%%%%%%%%%%%%%%%%%%%%

%%%%%%%%%%%%%%%%% BODY OF PAPER %%%%%%%%%%%%%%%%%%
\section{Introduction}
\label{introduction}

S0 galaxies feature a large central bulge surrounded by a smooth disk with no spiral arms, consist mainly of old, red stellar populations and contain very little gas. They make up a significant fraction of all galaxies in the local Universe, particularly in high-density cluster environments \citep{1980ApJ...236..351D}. Yet despite their ubiquity, significant uncertainty remains as to how these galaxies formed.

The recent advent of large scale surveys, in particular integrated field unit (IFU) spectroscopic surveys which allow for the creation of resolved 2D maps of various galaxy properties, has allowed for an unprecedented view of a large number of galaxies through many different environments, including S0s \citep{2018MNRAS.481.2299S,2012A&A...538A...8S}. Large scale cosmological simulations can now be used to reproduce the observational properties of galaxies in these surveys and to directly trace their evolutionary histories back in time, allowing for investigations into which formation pathways lead to the S0 galaxies we observe today.

\begin{figure*}
\includegraphics[width=2\columnwidth]{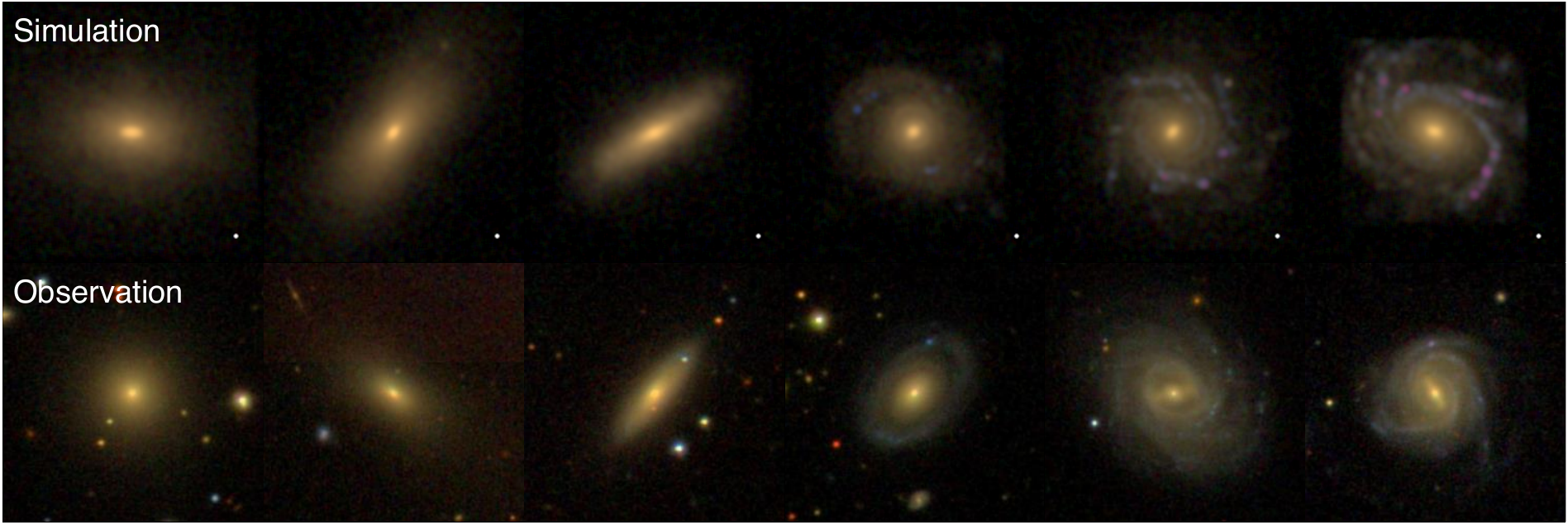}
   \caption{Comparison of mock rgb SDSS images generated from IllustrisTNG (top row) with real SDSS images of equivalent SAMI galaxies (bottom row). The mock SDSS images were used to classify the galaxies.}
 \label{clustering}
\end{figure*}

The proposed formation pathways of S0 galaxies generally fall into two main categories. The first major category involves the stripping of gas from a blue star-forming spiral galaxy as it falls into a dense environment, usually referred to as ram pressure stripping \citep{2000Sci...288.1617Q,2005AJ....130...65C}. This occurs when a spiral galaxy falls into a large, dense environment such as a massive galaxy group or cluster, where its internal gas interacts with and gets forced out by the dense intra-cluster medium. Since the disk structure of the spiral is left relatively undisturbed by this process, the final S0 is expected to retain a low disk-like S\'ersic index and a high degree of rotational support (though it has been shown that the group environment can result in an increase in size of the central bulge, which would raise the overall S\'ersic index \citep{2011MNRAS.415.1783B}. Galaxies actively undergoing this process have been observed in many clusters and are commonly referred to as Jellyfish galaxies \citep{2014ApJ...781L..40E}. 

The second major group of pathways involves a more disruptive merger event \citep{2017A&A...604A.105T,2015A&A...573A..78Q}. In this process, it is believed that a merger between a star-forming spiral and a smaller galaxy disrupts the internal gas within the spiral galaxy, either causing it to be flung outwards or consumed in a burst of star formation. Since merger events are generally more common in smaller groups relative to clusters (due to the increased relative speeds and therefore shorter interaction times between cluster galaxies), this pathway may be expected to occur more frequently in the lower density environments. Fiducial simulations have shown S0s can form from a spiral-spiral merger \citep[for example][]{1998ApJ...502L.133B}, and more recently \citet{2020MNRAS.493.1375P} used IllustrisTNG to demonstrate that disk galaxies can form from major merger events in a cosmological context. Such events are expected to increase the concentration of the galaxy and decrease its angular momentum \citep{2017IAUS..321..114Q}. 

We have also proposed an additional pathway, where an established compact elliptical galaxy is able to re-form an internal disk and transform into an S0 \citep{2018MNRAS.477.2030D}. In this scenario, the elliptical galaxy experiences a merger with a small gas-rich satellite galaxy. The gas and stars of the smaller galaxy collapse into a disk around the elliptical, accompanied by a burst of star formation within the accreted gas. This process results in an old central core and a young blue disk, which eventually consumes the gas via star formation and grows red. However, the fiducial simulation was set up in isolation and tuned to try and produce an S0; whether this process can occur naturally in the Universe remains unanswered. 

Recent observational work has suggested that multiple pathways may be active \citep{2020MNRAS.498.2372D,2018MNRAS.481.5580F,2020MNRAS.495.1321D}. A recent study of over 200 S0s using the Sydney-AAO Multi-object Integral-field Spectrograph (SAMI) galaxy survey \citep{2020MNRAS.498.2372D} showed that S0s cover a wide range of the kinematic and structural parameter spaces, which is difficult to explain with a single formation pathway. Instead, it was argued that S0s with lower degrees of rotational support are formed via disruptive merger events, while those of higher rotational support were formed from faded spiral pathways. \citet{2020MNRAS.492.2955C} found that S0s in field environments have lower rotational support relative to cluster S0s, suggesting that field S0s originate via mergers and cluster S0s originate via a rapid removal of gas. Comparisons of passive (red) spirals with S0s in CALIFA \citep{2019ApJ...880..149P} also showed that some S0s are consistent with a spiral origin while others are more in line with a merger origin.  The luminosities of many S0s appear to be too bright for them to have originated directly from a spiral \citep{2005ApJ...621..246B}. It is also difficult to explain the existence of S0s in a large range of environments using a single pathway, since mergers are less common in clusters and ram pressure stripping cannot occur in small groups, where S0 formation is surprisingly more prevalent \citep{2009ApJ...697L.137P}. Observations of young bulges and older disks in cluster S0s \citep{2012MNRAS.422.2590J} and old bulges with younger disks in isolated S0s \citep{2017MNRAS.466.2024T} further supports the occurrence of different formation histories within different environments, and a study over all galaxy types found signs of both inside-out and outside-in quenching \citep{2019ApJ...872...50L}.  

Star forming rings have previously been observed in S0s and early-type spirals. These have often been attributed to fresh gas infalls \citep{2019AJ....158....5P} or a merger with a small gas-rich galaxy which has already transformed into an S0 \citep{2014MNRAS.439..334I,2015A&A...575A..16M}, rather than part of the transformation process itself. 

In addition to these recent large-scale observational surveys, new hydrodynamic simulations have allowed for unprecedented investigations into the formation histories of galaxies throughout the Universe. Such simulations have been used to investigate the origin of fast and slow-rotating ellipticals \citep{2017MNRAS.468.3883P}, the quenching of galaxies \citep{2020MNRAS.tmp.2921D,2020MNRAS.496.2673J}, and the remnants of mergers \citep{2020MNRAS.493.3716H}.  The data produced from these simulations also allow for the derivation of present-day observable such as kinematic maps \citep{2019MNRAS.487.2354B} and optical images \citep{2015MNRAS.454.1886S} in such as way as to match what is currently achievable with observational surveys. Combining these simulations with the observational surveys presents a new opportunity to further investigate the formation histories of S0 galaxies. 

In this work, our aim is firstly to determine if both the merger and faded-spiral pathways are occurring, along with their relative contributions to the final population. Secondly, we aim to confirm if the new merger pathway proposed in \citet{2018MNRAS.477.2030D} can occur in a cosmological context, and if so, to determine its relative importance. To achieve these aims, we use the extensive data generated from the IllustrisTNG simulation to replicate present-day observations of SAMI galaxies and identify S0 galaxies, and then take advantage of the full histories recorded in the simulation to trace these galaxies back through time, identifying how they formed and what events they have experienced. We then link these different histories to the distributions observed in SAMI and argue that this distribution is the result of multiple formation pathways. The simulation used in this work assumes a ${\rm \Lambda CDM}$ cosmology with ${\rm \Omega_{m}}=0.3089$, ${\rm \Omega_{\lambda}}=0.6911$ and ${h=0.6774}$.

\section{Data}
\label{data}

\subsection{The IllustrisTNG-100 Simulation}

In this work we use the IllustrisTNG-100-1 simulation \citep{2019ComAC...6....2N,2018MNRAS.480.5113M, 2018MNRAS.477.1206N, 2018MNRAS.475..624N, 2018MNRAS.475..648P,2018MNRAS.475..676S}. The simulation was carried out over a co-moving volume of $106.5 ~\rm Mpc^{3}$ with a resolution of $9.4\times10^{5}$ and $5.1\times10^{6}~\rm M_{\odot}$ for baryonic and dark matter particles respectively. IllustrisTNG-100 consists of large scale structures up to a mass of $10^{14.3} ~\rm M_{\odot}$ (corresponding to a Virgo-sized cluster). This structure mass range covers the mass range of the SAMI galaxy survey, however it doesn't contain the high mass structures contained within the SAMI cluster sample ($~10^{15} ~\rm M_{\odot}$). We chose the IllustrisTNG-100 run due to its balance between the simulated volume and the resolution achieved - this offers us the chance to investigate a wide range of environments up to Virgo-sized halos, while maintaining a high level of resolution capable of clearly resolving the bulge and disk structures as well as the gas distributions. Unfortunately this means we are unable to include high-mass supercluster environments, which should be kept in mind throughout this paper. Due to this limitation, all comparisons with observational data are made using only the group-dominated Galaxy and Mass Assembly (GAMA) survey fields of the SAMI survey, leaving out the eight cluster fields. 

In this work we make use of various catalogues available for the IllustrisTNG simulation. We use the group catalogue for the masses of each galaxy's host group halo, and the Subfind Subhalo catalogue for the properties of individual galaxies such as stellar and gas masses, star formation rates and position. To determine how many stars are in a disk-like orbit, we use the Stellar Circularities catalogue \citep{2015ApJ...804L..40G}. We use the Stellar Projected Sizes catalogue for the half-light radius, as measured for a random orientation (i.e. always along the z-axis) \citep{2018MNRAS.474.3976G}. Stellar photometries by \citet{2018MNRAS.475..624N} are used to determine the $g-r$ colour index. To create the mock red-green-blue optical images we combine the idealised band images created by \citet{2019MNRAS.483.4140R}. Finally, we use the Sublink merger trees to identify merger events in each galaxy's history \citep{2015MNRAS.449...49R}. 

\subsection{Galaxy samples}

In order to allow for a direct comparison with previous observational work, we extracted a random sample of 500 galaxies from the final snapshot of IllustrisTNG-100 mass-matched to the galaxy sample in the GAMA region of the SAMI survey \citep[using the SAMI Data Release 2,][]{2018MNRAS.481.2299S}. We imposed an additional lower stellar mass limit of $10^{9.5}~\rm M_{\odot}$ in order to ensure any disk components can be clearly resolved in all galaxies. The random selection was carried out over the entire simulation volume, covering all environments. The methods used to visually identify our final sample of S0 galaxies in IllustrisTNG are detailed in Section~\ref{galaxy sample}. 

\section{Methods}
\label{methods}

In this section we describe how we classified the galaxy sample, derived the observationally-equivalent  physical parameters and analysed the formation histories of galaxies in Illustris.

\subsection{Morphological Classification}
\label{galaxy sample}

To allow for direct comparisons with our observational results, it is critical to ensure that we are looking at galaxies in the simulation which correspond to those classed as S0 in the SAMI survey. To this end, we looked to follow the same visual classification scheme used for the SAMI survey.

\begin{figure}
\includegraphics[width=0.9\columnwidth]{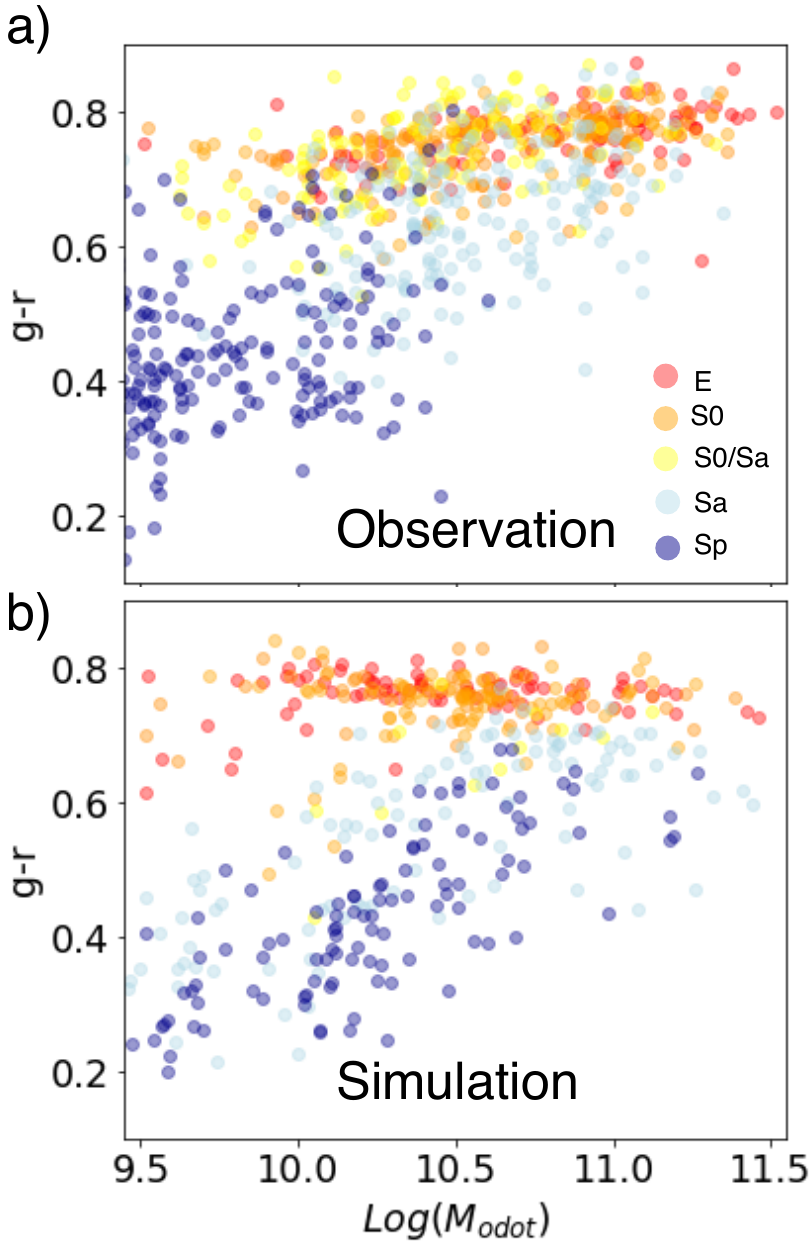}
   \caption{Comparison of our IllustrisTNG sample with the SAMI survey in the colour-mass plane. Both samples are coloured by their visual classifications. The red sequence in the simulation has a negative slope compared to the positive slope in the observations (as has been previously noted, see text) and different relative distributions of early and late type spirals. Overall however, the distributions and classifications of the simulation sample broadly follow the observational sample.}
 \label{samples}
\end{figure}

\begin{figure}
\includegraphics[width=0.9\columnwidth]{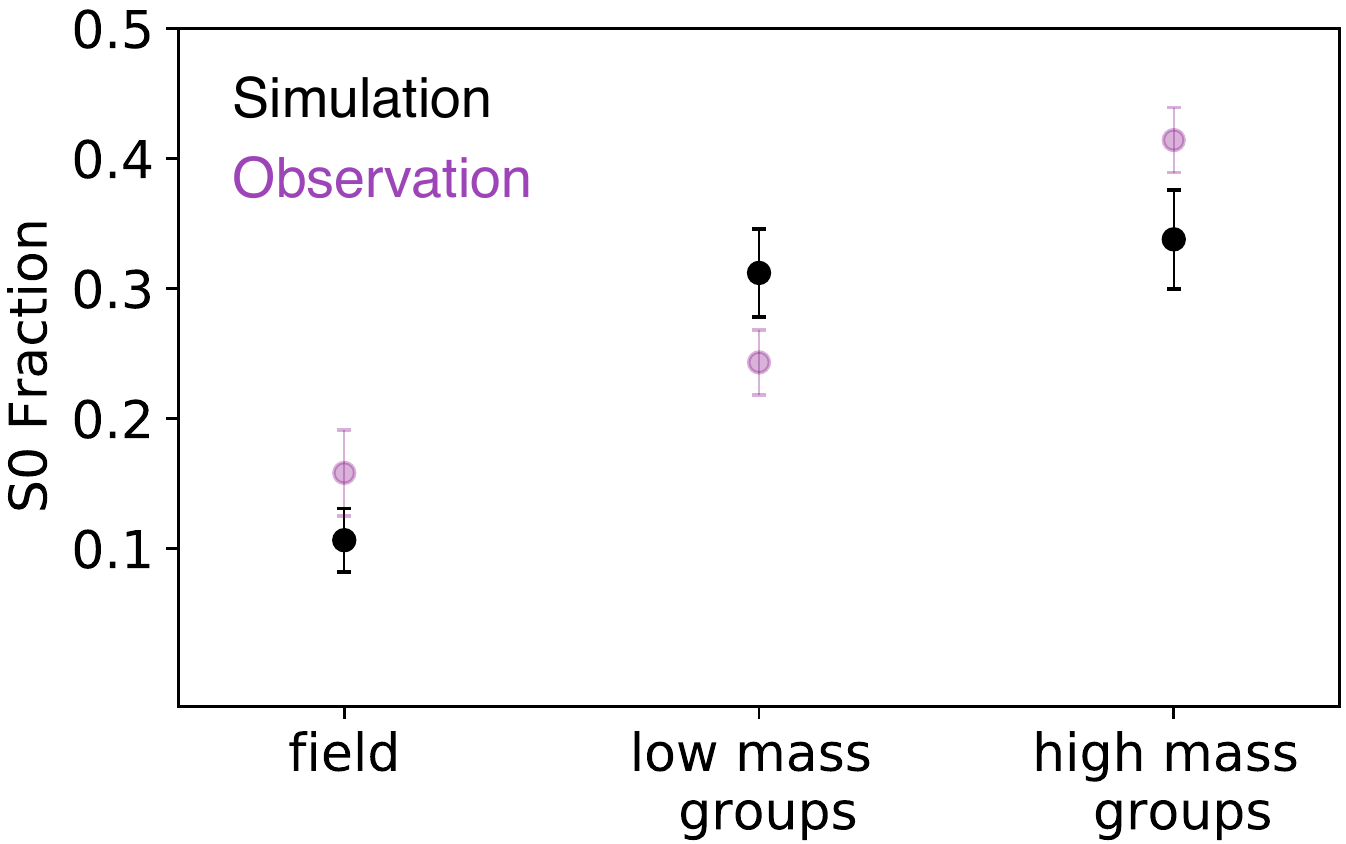}
   \caption{Fraction of galaxies classed as S0 as a function of their host environment for both our IllustrisTNG-100 simulation sample (black) and the SAMI observational sample (purple). Field, low mass groups and high mass groups correspond to host halos with a total mass less than $10^{12}~\rm M_\odot$, between $10^{12}$ and $10^{13}~\rm M_\odot$, and above $10^{13} ~\rm  M_\odot$ respectively. The relative fraction of S0s in our simulation sample closely follow those of observations across different environments. The cluster fields in the SAMI survey are left out due to the lack of these environments within the simulated volume.}
 \label{group_frac}
\end{figure}

\begin{figure*}
\includegraphics[width=1.7\columnwidth]{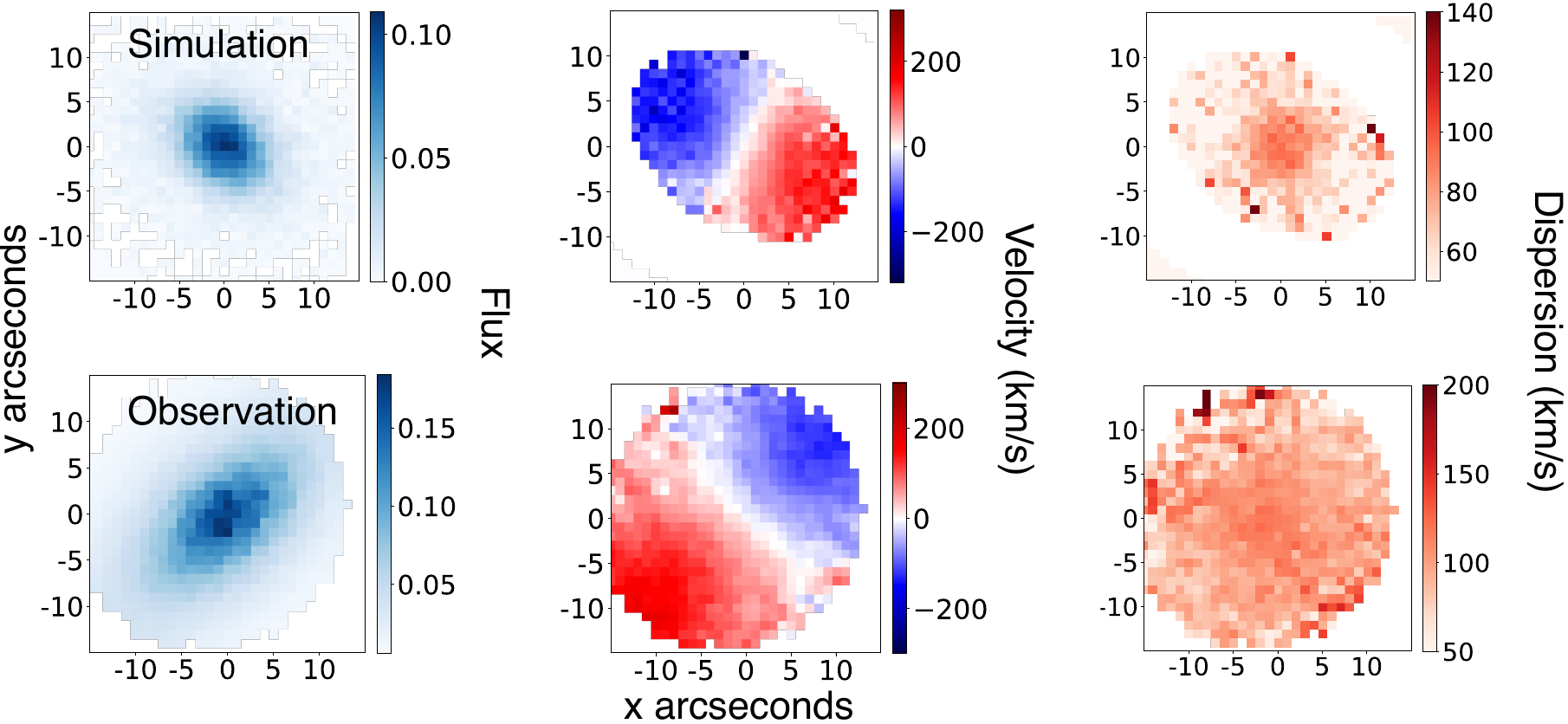}
   \caption{An example of the stellar kinematic maps derived for the IllustrisTNG-100 galaxies (top row) compared to those of an example galaxy from SAMI (bottom row). The left column shows the flux, the middle column shows the line-of-sight velocity map, and the right-most column displays the velocity dispersion map.}
 \label{clustering}
\end{figure*}

We generated mock Sloan Digital Sky Survey (SDSS) images which resembled those used in the visual classification. Idealised images in the $g$, $r$ and $i$ SDSS bands were generated by \citet{2019MNRAS.483.4140R} for IllustrisTNG galaxies using the radiative transfer code SKIRT \citep{2015A&C.....9...20C}. Firstly, we adjusted the angular sizes of the images to match the range in angular sizes present in the SAMI sample. We restricted the minimum size to 2" per effective radius ($\rm Re$), so that all galaxies were large enough to identify key components such as the bulge, disk and/or spiral arms. This was done to minimise the number of galaxies flagged as 'unknown' and to provide us with a larger final sample from the given set of classification images. The idealised images were then convolved with the SDSS point spread function (1.31 arc-seconds) and then combined to form a red-green-blue colour image, following \citet{2015MNRAS.454.1886S}. We then applied a random Gaussian noise corresponding to a sky signal-to-noise ratio of 1/10. The resulting 500 images were visually classified by six of the authors, following a similar scheme to that used in SAMI; Firstly, the galaxies were identified as either early-type or late-type. Secondly, early-type galaxies were further separated into those which featured a distinct disk component and those which did not, while late-type galaxies were further separated into those dominated by spiral arms and those with weak spiral arms. This last step was altered from the original scheme, as IllustrisTNG tends to produce galaxies with a larger spheroidal to total mass ratio, particularly for low-mass galaxies where disc-only structures are more prominent in SAMI \citep{2019MNRAS.487.5416T}. 

The results of the classification are compared to the SAMI visual classifications in Figure~\ref{samples}. The IllustrisTNG classifications reproduce the blue cloud and red sequence in SAMI, demonstrating that our classifications are reasonable. The red sequence has a different slope in illustrisTNG relative to SAMI, a previously known feature in cosmological simulations likely due to a difference in the mass-metallicity relation between IllustrisTNG and observations \citep{2018MNRAS.475..624N}. In addition, the distributions of early and late type spirals are more similar to each other in our illustrisTNG sample compared to SAMI, where they are more clearly separated (due to the lack of bulge-less galaxies as mentioned above). As a test of our classification approach, Figure~\ref{group_frac} displays the fraction of galaxies classed as S0 as a function of environment. For the observed sample we use the group masses in the Galaxy and Mass Assembly group catalogue \citet{2011MNRAS.416.2640R}. The environmental dependence of the S0 fraction in IllustrisTNG closely matches SAMI, which is a further validation that our classification approach produces realistic results and is consistent with SAMI. 

\subsection{Kinematics}

Following \citet{2019MNRAS.487.2354B}, in order to allow for direct comparisons of S0 kinematics between the SAMI observations and IllustrisTNG, we derived kinematic maps which resembled those created in SAMI. Firstly, for each galaxy, we extracted all stellar particles associated with its halo and mapped them in the x-y plane, along with their velocities along the z-axis which is taken here to be the line of sight. Over this plane we overlaid a grid with a cell size equal to the spatial scale of the SAMI survey maps (around 0.492 kpc per spaxel at the median SAMI redshift of $z=0.1$). For each cell, we created a Gaussian distribution with a FWHM equivalent to the SAMI seeing half-width of $1.31"$, and created a histogram of stellar velocities weighted by that distribution. The histogram was re-binned to a resolution of 8.8 km/s, representing the spectroscopic resolution of SAMI. A separate Gaussian function was fitted to this histogram, and the mean of the histogram was set equal to the line-of-sight velocity of that cell, while the FWHM was used to derive the dispersion of that cell. Finally, the SAMI flux and kinematic maps were used to derive the relationships between the flux and the flux and velocity uncertainty; this was then used to apply a flux-dependent noise pattern to the kinematic maps

To assess the degree of rotational support, following \citet{2020MNRAS.498.2372D} we employed the widely-used parameter $v/\sigma$, where v is the observed line-of-sight velocity and $\sigma$ is the velocity dispersion. We calculate $v/\sigma$ using the following equation:

\begin{figure*}
\includegraphics[width=2\columnwidth]{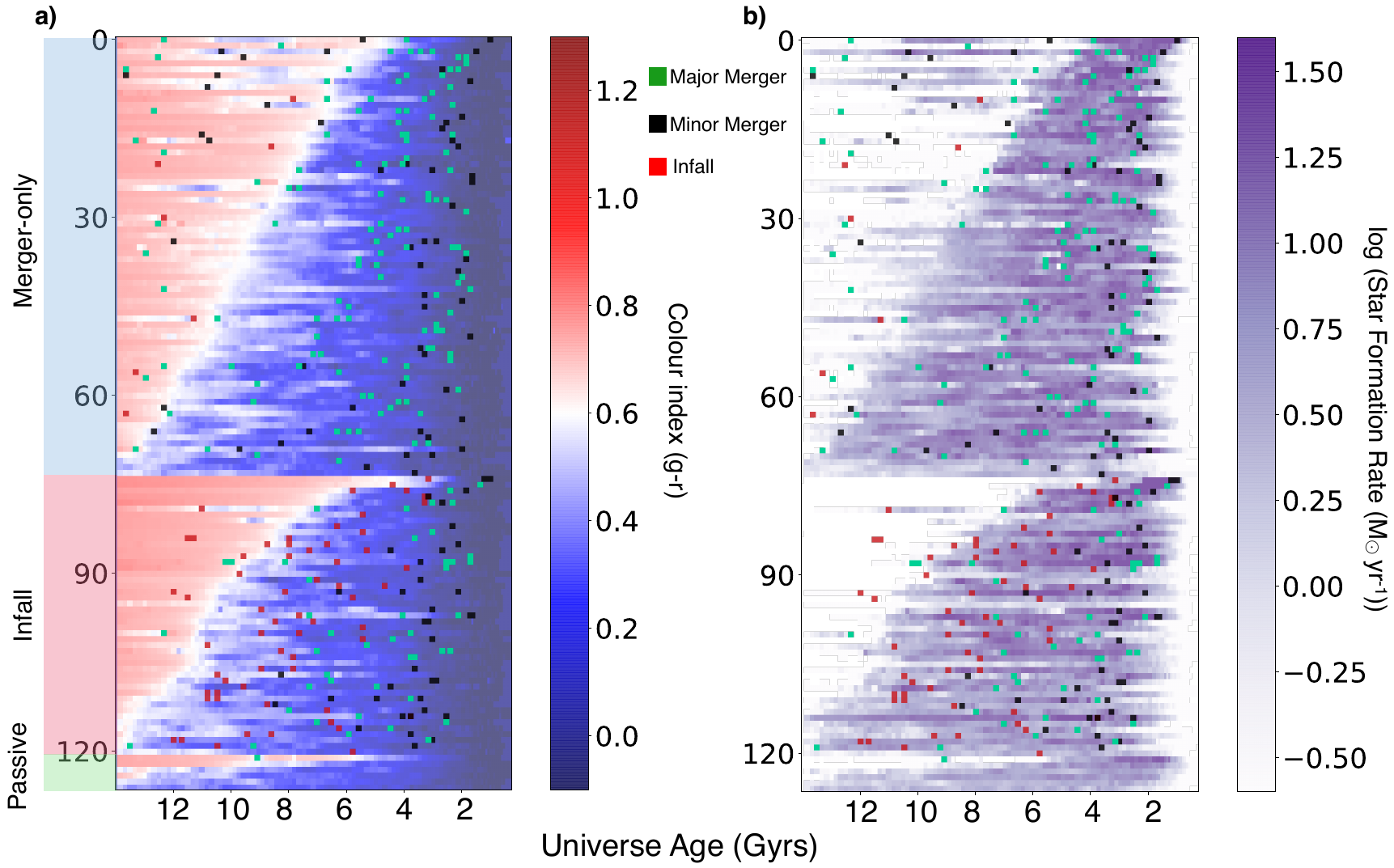}
   \caption{These figures display the complete evolutionary history of every S0 galaxy in our simulation sample. Shown here is the evolution of colour (a) and star formation (b), along with all significant events they have experienced during their formation. Marked on the plots are group infall events (red points), minor mergers (blue points) and major mergers (green points). Galaxies are placed into three groups according to the last significant event before transformation (as labelled on the left axis of a), and within these groups they are arranged by the time at which they cross the g-r colour index of 0.6 while transitioning from blue to red.}
 \label{histories}
\end{figure*}

\begin{equation}
\left(\frac{v}{\sigma}\right)^{2} = \frac{\sum_{i=0}^{N_{spx}}{F_{i}V_{i}^{2}}}{\sum_{i=0}^{N_{spx}}{F_{i}\sigma_{i}^{2}}},
\label{refinement_eq}	
\end{equation}

where the sum is over all spaxels within a de-projected radius of 1.5 ${\rm R_{e}}$ and $F_{i}$ is the flux of spaxel $i$. Values of $v/\sigma$ approaching 1 indicate a rotationally supported structure, while values towards zero indicate a pressure-supported structure. The de-projected radius of each spaxel corresponds to the major axis of the ellipse on which it is located, which was determined using the $r$-band-derived axis ratio and position angle.

\subsection{Identifying Formation Pathways}

In following the history of each S0 in our sample and identifying their respective formation pathways, we focused on two events; a merger event where a secondary halo merges with the main progenitor, and an infall event where the galaxy falls into a significantly denser environment. 

To search for merger events, we use the LHaloTrees catalogue \citep{2005Natur.435..629S} and follow the galaxy's progenitors backwards through time. To find merger events, we look for a new secondary progenitor appearing along side the primary galaxy (i.e. when the 'Next Progenitor' flag is non-zero), which must have merged into the primary galaxy before the following snapshot. This following snapshot is then used as the time of the merger.

 For this work, we used the stellar masses of the primary and secondary halo to calculate their mass ratios since this can be more directly linked to observations. A well-known issue in merger trees is the flow of particles from the secondary to the primary halo before the merger occurs; if the merger ratio is calculated at the point of merger, the ratio can be significantly underestimated. We therefore trace both the primary halo and the secondary halo back 20 snapshots (or up to the beginning of their existence, if this is less than 20 snapshots) and identify the point at which the secondary is at its greatest mass. The mass of the primary and secondary halo at this snapshot is then used to calculate the merger ratio. In addition, to counteract the known halo switching problem \citep[where close halos can be swapped around and incorrectly assigned to the primary or secondary galaxy, see for example][]{2017MNRAS.472.3659P} and keep the merger ratio below 1, we looked for switches in the primary and secondary halo masses during the merger and swapped the masses around when this occurred. The merger event is flagged as a minor merger if the stellar mass ratio is greater than 1:10 and less than 1:3, or a major merger if the mass ratio is greater than 1:3.

\begin{figure*}
\includegraphics[width=2\columnwidth]{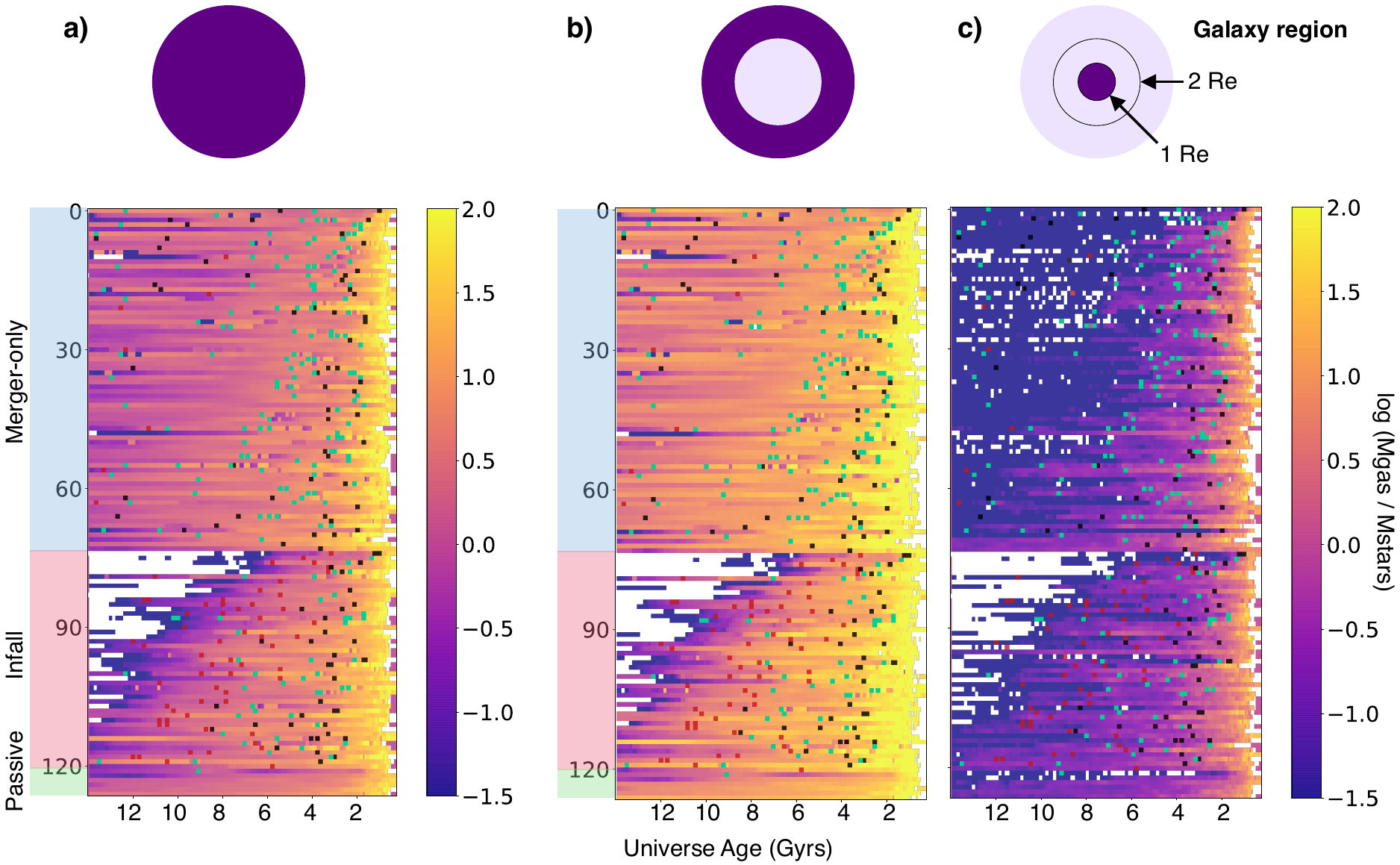}
   \caption{History plots of the gas contents of S0 galaxies, within the galaxy regions highlighted by the dark purple regions in the circles along the top. a) shows the total gas mass normalised by the stellar mass, highlighting the stripping occurring for the passive and infall pathways. b) and c) show the gas content within 1 Re and beyond 2 Re respectively, illustrating that while the merger group retain a large amount of gas, this gas is located in the outskirts while the central regions have become largely devoid of gas.}
 \label{gas}
 \end{figure*}

To identify infall events, where a galaxy falls into a significantly larger group or cluster, we tracked the total mass of the host group halo for each galaxy through time and flagged any events where this mass suddenly increased by half an order of magnitude. Such an increase occurs between subsequent snapshots when the friends-of-friends group finding algorithm assigns the galaxy to the larger group in the following snapshot. Variations of this threshold did not significantly affect the results. 

\section{Results}

Firstly, we show the formation histories of our IllustrisTNG S0 sample, including the significant events each galaxy experienced, and argue that these histories fall into two main groups. We then look at the gas evolution and merger trees of S0s in each group, further supporting the claim of two different pathways. We show examples of each pathway, as well as comparisons of star-forming ring structures with SAMI $\rm H\alpha$ maps. We then show an example of an S0 forming from an elliptical, and finally we look at red spiral galaxies and compare these to the S0 formation histories. 

\subsection{Galaxy histories in IllustrisTNG}

We firstly looked to identify which formation pathways are active in our S0 sample and to place them into groups according to their pathway. Figure~\ref{histories} displays the history of all S0 galaxies in our IllustrisTNG sample. Here, as an approximation of when the transition to an S0 took place, we use the time at which the galaxy's $g-r$ colour index first crossed 0.6 as it changes from blue to red (this value was chosen using the colour index distributions of S0s and spirals seen in the observational sample, see Figure 2). Visual inspection of stellar luminosity images (such as those presented in Section~\ref{stripped}) indicated that in most cases, the loss of spiral structure indeed corresponds to when the galaxy crosses this colour threshold. We flagged the following significant events which occurred in each galaxy's history; minor mergers (with a mass ratio of between 0.1 and 0.3), major mergers (with a mass ratio greater than 0.3) and infall events (where the total mass of the host group halo increased by at least a factor of five, indicating that the galaxy had become gravitationally bound to a larger group or cluster environment). We grouped the S0s by the events they have experienced before the transformation occurred, and within each group, ordered them by their transformation time. The two major groups we identified were those which had experienced only a significant merger event before the transformation, and those that had experienced an infall into a more massive environment. 

Thirteen galaxies were found to have no significant events flagged before the transformation occurred. The merger histories of these galaxies (investigated using figures such as those in Section~\ref{merg}) revealed that some of these could also be placed into one of the above two groups; four of these galaxies became red during an extended interaction with another galaxy which subsequently merged after the transformation, meaning that the interaction and merger process still caused the transformation even though the flagged time of the merger event occurred afterwards. These galaxies were therefore placed into the first major group identified above. Three galaxies fell through a large group without becoming gravitationally bound to the group prior to the transformation, and followed an evolution consistent with others in the second group. The remaining six galaxies have no sign of any significant merger activity or infall events, and are placed into a separate group we label as 'Passive', shown at the bottom of Figure~\ref{histories}. 

Figure~\ref{histories} also illustrates that the transformation rate of S0s has been occurring relatively constantly over much of the Universe's history, from 10 Gyrs ago up to the present day. The time taken to cross the colour boundary is also quite rapid in most cases, though a few galaxies go through subsequent periods of star formation before finally settling into a passive phase.

\begin{figure*}
\includegraphics[width=2\columnwidth]{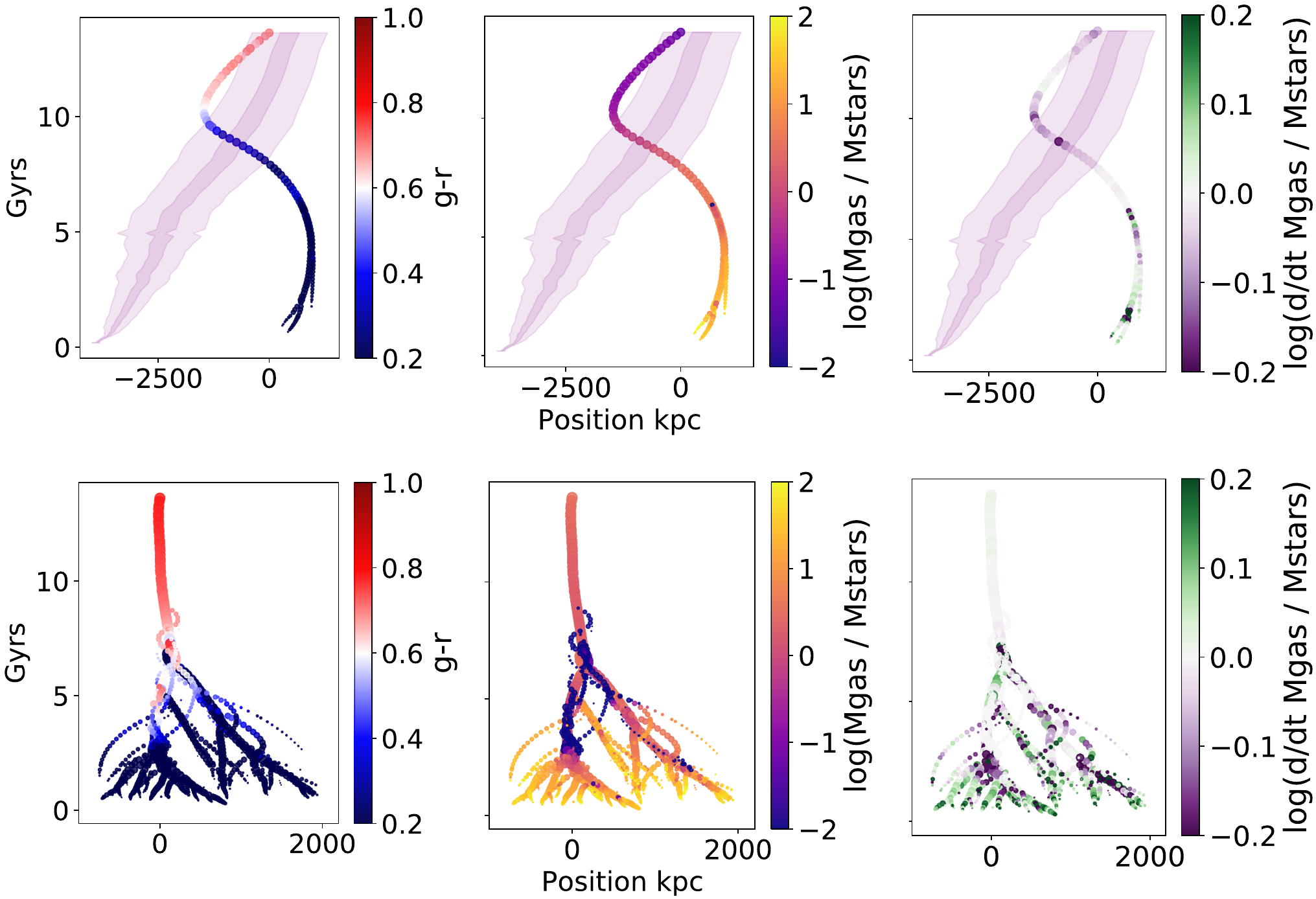}
   \caption{Typical examples of position-time plots (projected onto the x-axis) for S0s forming via a group infall event (top) and a major merger event (bottom). The position and stellar mass of the progenitor galaxy, along with every other galaxy which merges into it, is shown at each snapshot by the location and size of the points respectively. The points in the left column are coloured by the g-r colour index, the middle column points are coloured by the stellar-mass-weighted gas mass, and the points in the right plots are coloured by the change in gas mass between snapshots. The top example has a dynamic track as it interacts and falls into the cluster, while the straight track and prominent accretion history evident in the bottom example, accompanied by a steady gas mass is significantly different. Both galaxies have similar track forms in the other two spatial dimensions.}
 \label{trees}
 \end{figure*}

Within the infall group, a clear feature is the proximity of group infall events to the time at which the transition occurred. This suggests that a change in environment has a more rapid and dramatic impact on star-forming galaxies compared to the merger events, with the galaxies becoming red within 1 or 2 billion years of falling into a larger group or cluster. The range of timescales evident between the infall and transformation is likely due to several factors such as the host group mass, intra-cluster gas density and the proximity of the galaxy's passage through the group to its centre. There is very little correlation between the merger activity and transformation time within this group, showing that for these galaxies, it is the infall events which are the main drivers of the transformation. 

To check if the merger distributions are indeed different between the two groups, we compared the distribution of the timing of merger events relative to the time of transformation for each group using a two-sided KS test. We find that the probability that they come from the same underlaying distribution is 1.6 percent. This indicates that the time distribution of mergers is indeed significantly different between the two groups, as would be expected if they are playing a major role in one group but not the other.

\subsection{Gas histories}
\label{gas_section}

The star formation activity in galaxies is fuelled by their internal gas, and therefore the removal of gas is expected to play a significant role in the transformation. We therefore looked to determine how the different histories identified above impact the gas content of these galaxies.

Figure~\ref{gas} a) displays the stellar-mass-normalised gas content history for all of the S0s. Immediately evident in this plot is the significant amount of stripping occurring in the top half of the diagram, corresponding to S0s which have either experienced no major events, or have fallen into a large group or cluster environment. Many of these S0s become fully depleted in their gas soon after the infall event, indicating rapid stripping. The S0s that only experienced mergers, however, retain a significant amount of gas up to the present day. For these galaxies, the question then becomes how does the star formation get shut down, while still having relatively high amounts of gas.

\begin{figure*}
\includegraphics[width=1.9\columnwidth]{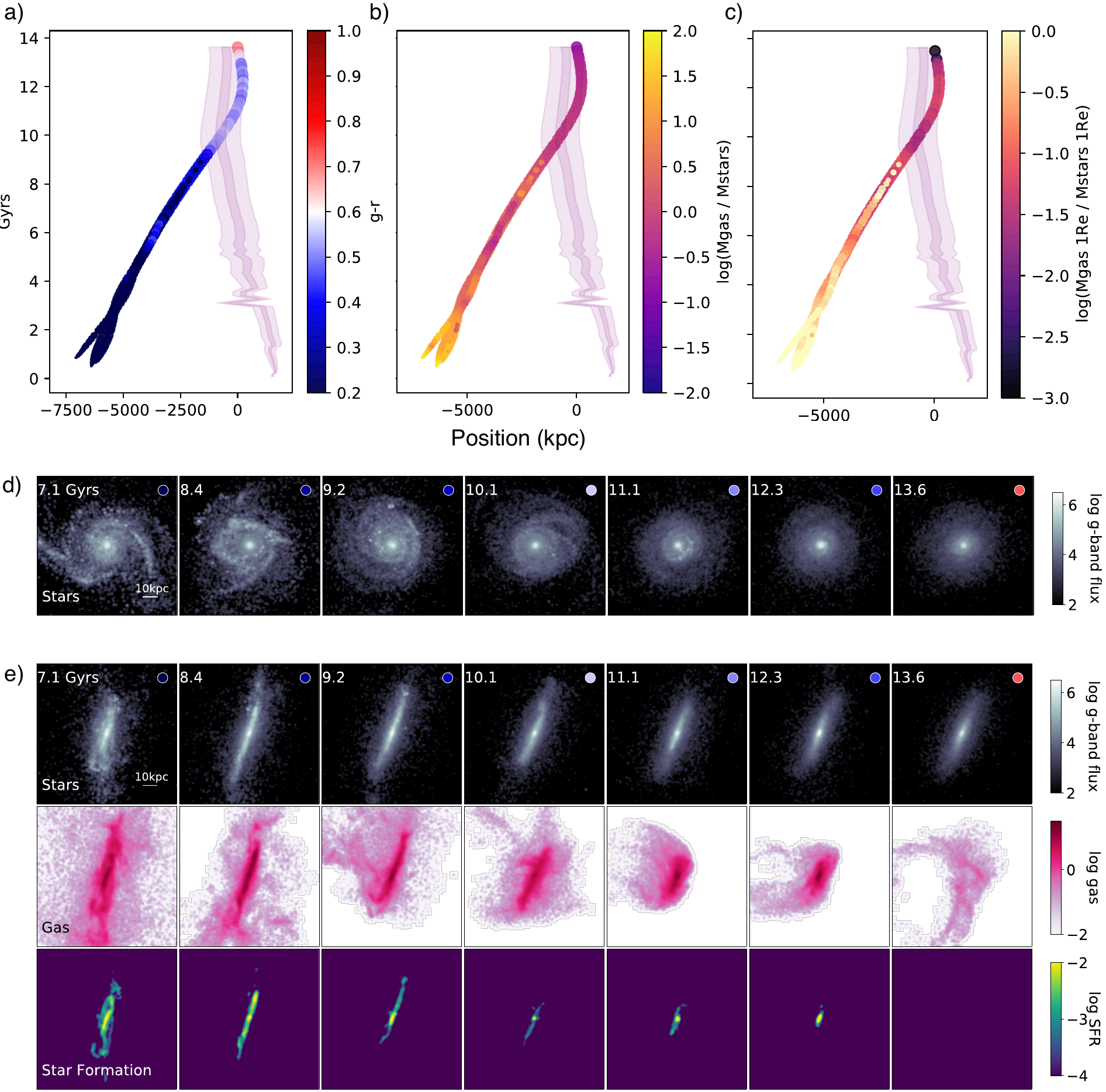}
   \caption{An example of an initially blue, gas-rich galaxy falling through the centre of a galaxy group and having its gas stripped out. The purple region in a), b) and c) shows the location and radius of the group in which the S0 is located in the final snapshot, followed back through time (using the central galaxy of this group). Light purple indicates the radius at which the local density exceeds 200 times the critical density (light colour) and dark purple shows $1/4$ of this radius. d) and the top row of e) show the face-on and edge-on projections of the $g$-band stellar luminosity respectively, at the times marked in the top left corners, as well as the galaxy's g-r colour index as indicated by the circle in the top right. The middle and bottom rows of e) display the distribution of the gas and star formation rates. The transition from blue to red begins when the galaxy passes through the centre of the host galaxy group. Initially the gas is concentrated in the central disc and accompanied by a large gas halo. This begins to be stripped out behind the galaxy at 9.2 Gyrs as the galaxy enters the group, and by the present day very little gas remains within the galaxy. This is accompanied by the disappearance of spiral arms and results in the smooth, featureless disk of an S0. }
 \label{jellyfish}
\end{figure*}

To further elucidate the differences between the merger-only and infall formation pathways, we then looked at the gas content as a function of radius. Figure~\ref{gas} b) and c) shows the total gas contained within 1 half-mass radius (1 Re) and outside of 2 Re respectively, normalised by the stellar mass. The gas histories of the stripped group have a similar appearance across all three plots, as would be expected from the complete and rapid removal of all gas. For the merger-only group, it is apparent in Figure~\ref{gas} b) that within 1 Re, most of the gas has been removed soon after a merger event, while outside of 2 Re they have experienced very little gas loss. This shows that while these S0s retain a significant amount of gas after the merger event, this gas is mostly located further out in the galaxy and its surrounding halo. This is in contrast to spiral galaxies which have experienced a merger event and are still forming stars today; in these galaxies, significant amounts of gas remain within 1 Re after the merger. 

Comparing Figure~\ref{gas} b) and c), the gas in the stripped population is getting depleted from the outside in, while the galaxies in the merger group are losing their gas from the inside out. Combining this with the observation of complete vs large-radii gas stripping discussed above confirms that the two groups identified here have very distinct evolutionary histories, further supporting the placing of S0s into two main groups.

\subsection{Merger histories}
\label{merg}

To further see what the differences are between the main groups of histories identified above, we then investigated the formation histories of individual galaxies using position-time plots. Figure~\ref{trees} shows typical examples of such plots for S0s with group infall events and major merger events, coloured by the galaxy colour, gas content and the rate of change of the gas mass.

\begin{figure*}
\includegraphics[width=1.9\columnwidth]{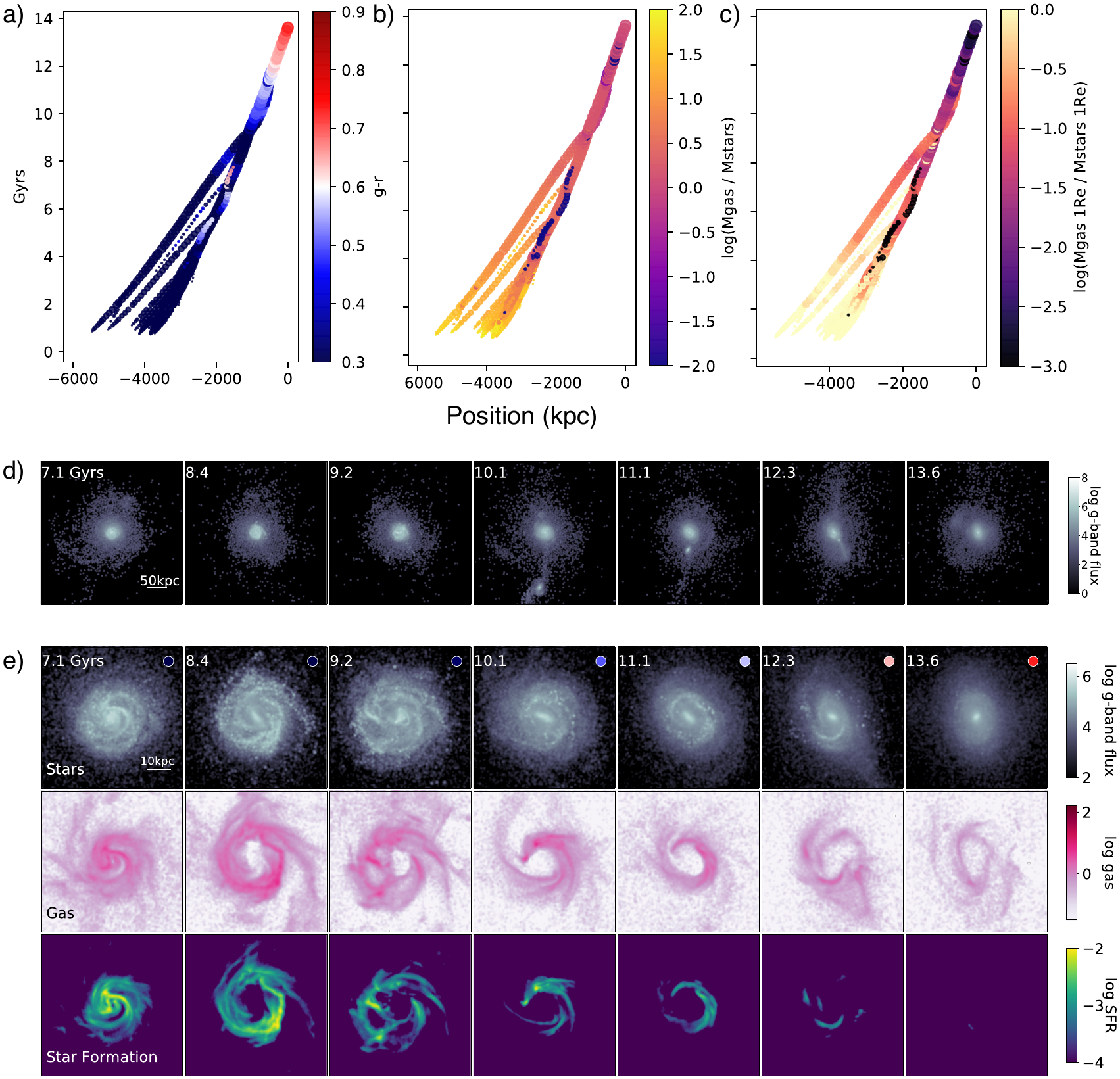}
   \caption{An example of a spiral galaxy transforming to an S0 after multiple merger events. a) and b) are the same as in Figure 8, while the points in c) are instead coloured by the gas within 1 effective radii to highlight the central gas depletion. d) shows the g-band stellar luminosity in a wider field of view to show a merging galaxy, while e) remains the same as in Figure 8. After the first significant minor merger (the last stage of which is visible as an over-density in the 1 o'clock position at 7.1 Gyrs in d), the gas within the central region has been depleted and a stellar bar has formed. As the remaining gas is consumed by star formation, it becomes concentrated into a narrow ring structure which then breaks up after 1-2 billion years, while the spiral arms begin to disappear. In this example, a second merging galaxy (visible from 10.1 Gyrs in d) then begins interacting with the progenitor and eventually merges, disrupting the remaining spiral arms. In examples where a second significant merger doesn't occur, the spiral arms still continue to weaken and eventually disappear, giving the final galaxy the morphology of an S0. }
 \label{ring_eg}
\end{figure*}

These plots again immediately show a significant difference between S0s which have experienced infall events and those which have had major merger events. Many in-falling S0s show very little accretion of mass after the initial buildup period, and dramatic turns in their motion. Looking at the rate of change of gas mass for the in-falling galaxies, the loss in gas typically begins when the galaxy enters the R200 radius of the group (where R200 is the radius at which the interior density is 200 times the critical density), indicating that it is indeed the infall event itself which is triggering the transformation. 

The trees of those undergoing merger activity typically show that, along with the significant event identified, they also experience more additional accretion of satellite galaxies, and tend to follow a very straight track through space. This suggests that these are more likely to be central galaxies through most of their history, while the previous tracks are indicative of either satellite galaxies, or transitions from central to satellite galaxies. 

\subsection{Evolution of stripped S0s}
\label{stripped}

We then further investigated S0s in the stripped group to see more clearly how and where they were losing all of their gas so rapidly. An example of a galaxy undergoing this process is presented in Figure~\ref{jellyfish}. To confirm that the rapid colour change of these galaxies was caused by the infall event, we followed the final host group of the S0 back through time relative to the S0's own path. We did this assuming that the central galaxy of this group remains the central galaxy back through time, and identifying the location and radius of this galaxy's host group at each snapshot.

\begin{figure*}
\includegraphics[width=1.6\columnwidth]{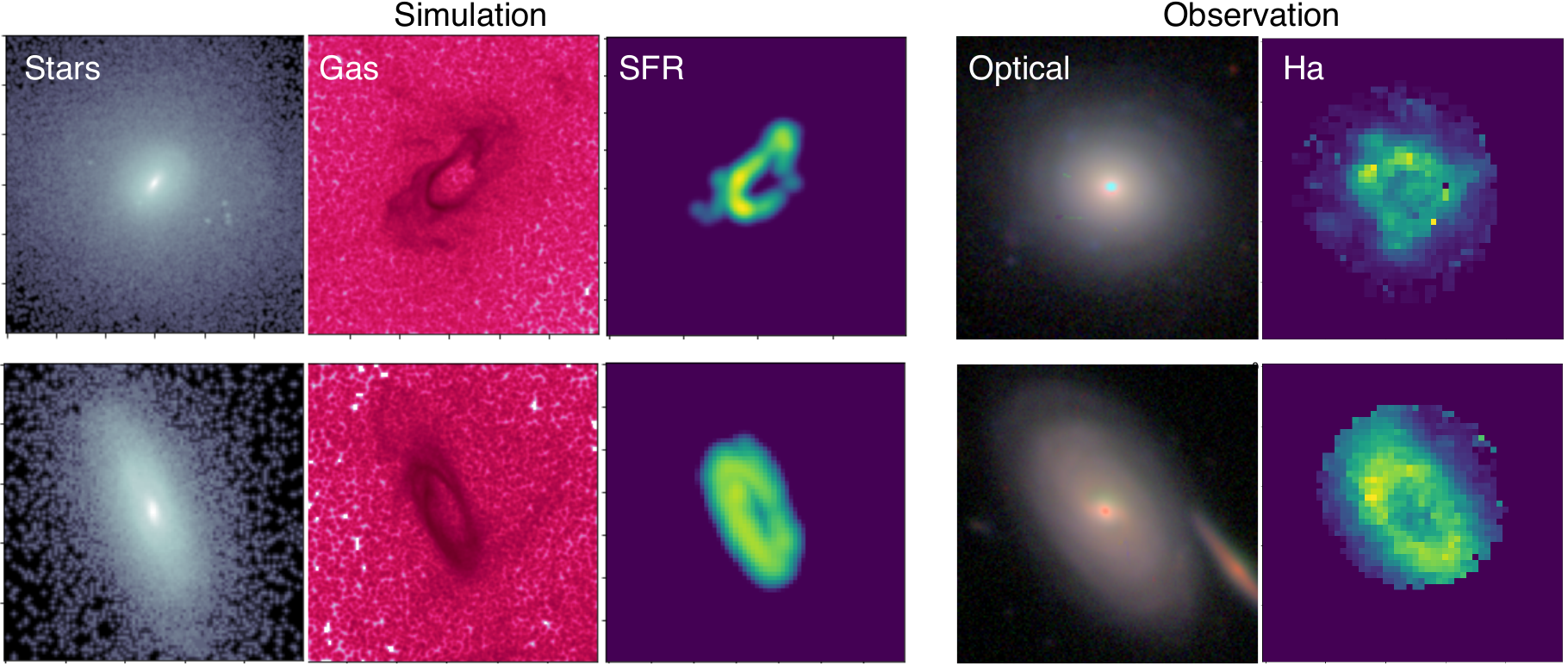}
   \caption{Comparison of the morphology of the star forming rings with $\rm H\alpha$ observations taken from SAMI. From left to right, the plots show distributions of stars, gas and star formation of two example illustrisTNG S0s, and the right column shows examples of the observed $\rm H\alpha$ flux in two SAMI galaxies. The first row illustrates a clumpy, irregular ring and the second illustrates a smooth, coherent ring, showing that both morphology types can be reproduced by the merger pathway.}
 \label{rings}
\end{figure*} 

This reveals that these galaxies transition from blue to red shortly after (or during) passing through the centres of the groups. As is evident in the example in Figure~\ref{jellyfish}, the colour change corresponds to a morphological change from a spiral structure to a smooth disk. This is highlighted particularly by looking at the rate of gas loss in relation to the group passage, where the gas loss begins right when entering the cluster's outer regions and peaks when passing the centre. Note that for this example plotted in one spatial dimension, the galaxy also passes through the centre in the other two spatial dimensions. While the galaxy is passing through the group centre, it features a filamented tail of gas trailing behind it, giving it the appearance of a jellyfish. Such jellyfish galaxies are frequently observed in dense cluster environments \citep{2012ApJ...750L..23O} and have been studied previously within Illustris \citep{2019MNRAS.483.1042Y}. (See Section~\ref{env} for a discussion on the environmental dependence of these pathways). Interestingly, the point at which the stripping begins almost always corresponds to when the galaxy passes through the R200 radius of the cluster; in some cases, the stripping pauses when the galaxy goes outside this distance on the other side, then resumes as soon as it re-crossed the R200 line. 

While this process of ram pressure stripping is well known and observed as 'jellyfish galaxies' \citep{2014ApJ...781L..40E}, the observations here place it into the broader context of the entire S0 population, and the dynamics described above (along with their occurrence frequency) can be directly compared to the merger group discussed below.

\subsection{Evolution of merger-only S0s}

As was revealed in Figure~\ref{gas}, the S0s in the merger group lose the majority of their gas from their central regions where the majority of stars are located, yet still appear to retain most of their extended gas envelope. In order to further understand this, we investigated in detail several galaxies from this group. A typical example of the stellar, gas and star formation rate evolution of one of these galaxies is presented in Figure~\ref{ring_eg}. 

After the initial build-up phase, the galaxy settles into a well-structured spiral galaxy with strong star formation activity within its nucleus and spiral arms. However, in the snapshot after the merger event, the central region has rapidly become depleted of gas, with much of the remaining gas condensed into the region outside the bulge in filaments and remanent spiral structures. The distribution becomes narrower and most notably forms a distinct ring structure with very low gas densities either side; this ring structure is visible for around 1-2 billion years. The star formation also becomes restricted to the ring. The formation of these rings may be caused by the gravitational interaction during the merger, resonances and/or a central bar. Indeed, looking at  Figure~\ref{ring_eg} we see the appearance of a bar-like feature when the central hole begins to open up (although in other cases, a bar is seen to from over a billion years before the formation of the ring). The formation of this bar may have been triggered by the interaction with the merging galaxy, which then subsequently removed gas from the central region. Investigations of the central black hole accretion rates before and after the merger events suggest that outflows from increased AGN activity are unlikely to have triggered the formation of these rings.

\subsection{The appearance of ring structures} 
 
Star formation activity is closely associated with $\rm H\alpha$ emission. Rings of $\rm H\alpha$ emission have been observed in S0s previously and are also present within the S0 sample of the SAMI galaxy survey. In Figure~\ref{rings}, we compare examples of the star formation rings found here with the $\rm H\alpha$ rings seen in SAMI S0s. The size, distribution and morphology of the SFR rings closely resemble the rings in SAMI, including both the rings with sharp turns, straight edges and significant clumps and smooth, continuous rings seen in other S0s.

\begin{figure*}
\includegraphics[width=1.9\columnwidth]{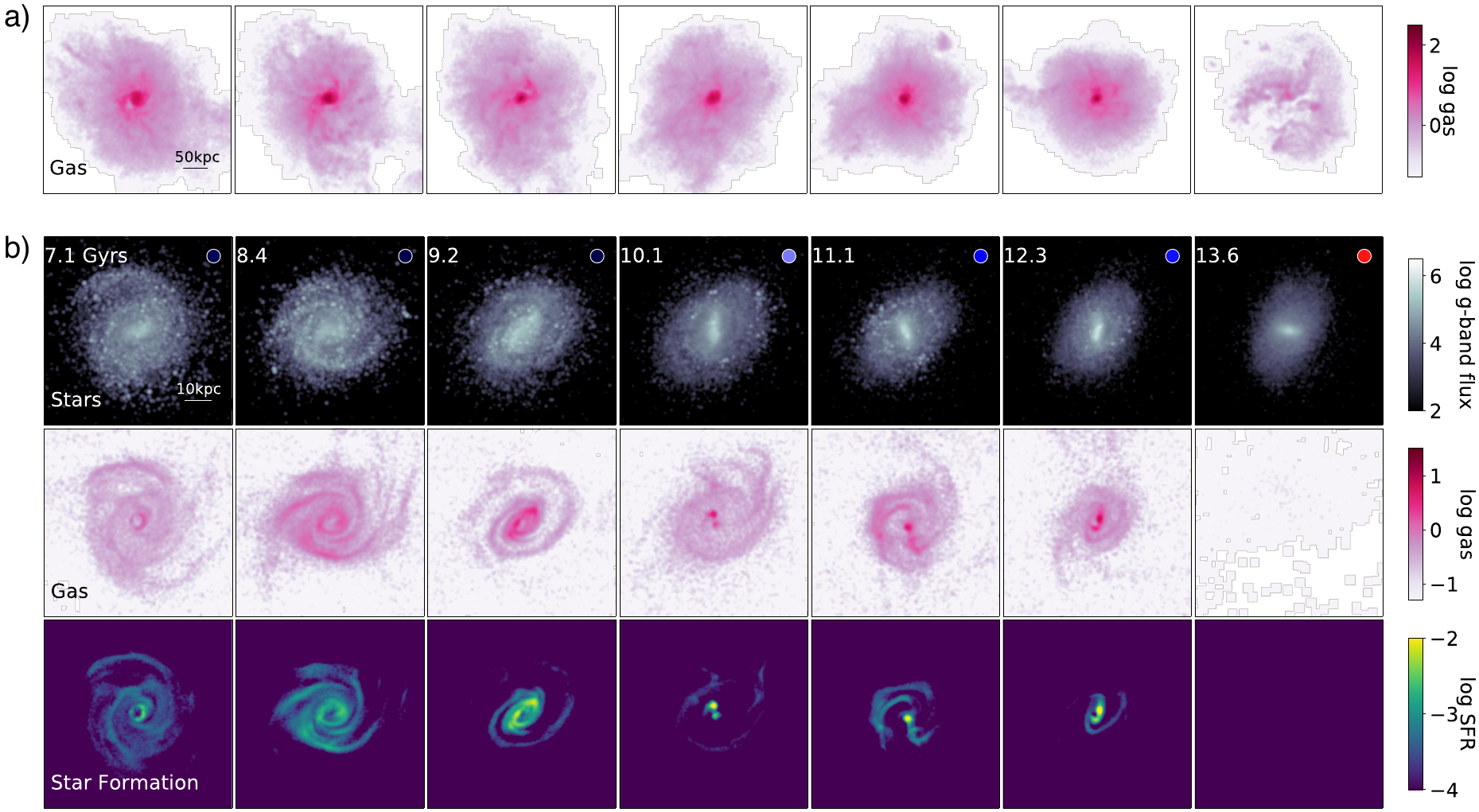}
   \caption{Typical example of the evolution of an S0 which experiences no significant events. a) shows a larger-scale view, highlighting the depletion of the extended gas halo. b) shows the evolution in the stellar luminosity, gas and star formation rate as in Figure 8, showing the outside-in quenching of star formation and the gradual loss of the spiral arms. }
 \label{isolated}
\end{figure*}

 Visual inspection of the SFR distributions of all S0s reveals that nearly all S0s in the merger group (and many in the infall group which experience a merger prior to the infall event) go through a phase where a clear SFR ring is present, appearing soon after the merger event. The average lifetime of the rings is very brief, typically around 1-2 billion years, meaning that despite their ubiquity, we would expect to observe only a small number of S0s with SFR rings in today's observations. Indeed, in SAMI we identify 10 clear rings out of 146 galaxies in the GAMA region, an occurrence rate of 7 percent. For comparison, the number of rings observed in snapshot 91 (at a redshift of 0.1) is 6.5 percent, remarkably close to the occurrence rate in SAMI. 

Figure~\ref{rings} compares the morphology of two galaxies during the transition to S0s to $\rm H\alpha$ rings observed in SAMI galaxies. The morphology of the rings produced in illustrisTNG closely resembles examples seen in SAMI, including both cases of clumpy and irregular rings seen in some galaxies as well as complete, continuous rings seen in others. This suggests that galaxies with observed $\rm H\alpha$ rings are actively undergoing the transformation from a blue spiral to a quiescent S0. This is in contrast to most previous suggestions that such rings are formed from an infall of fresh gas onto the S0 \citep[e.g][]{2015A&A...575A..16M}

\begin{figure}
\includegraphics[width=0.9\columnwidth]{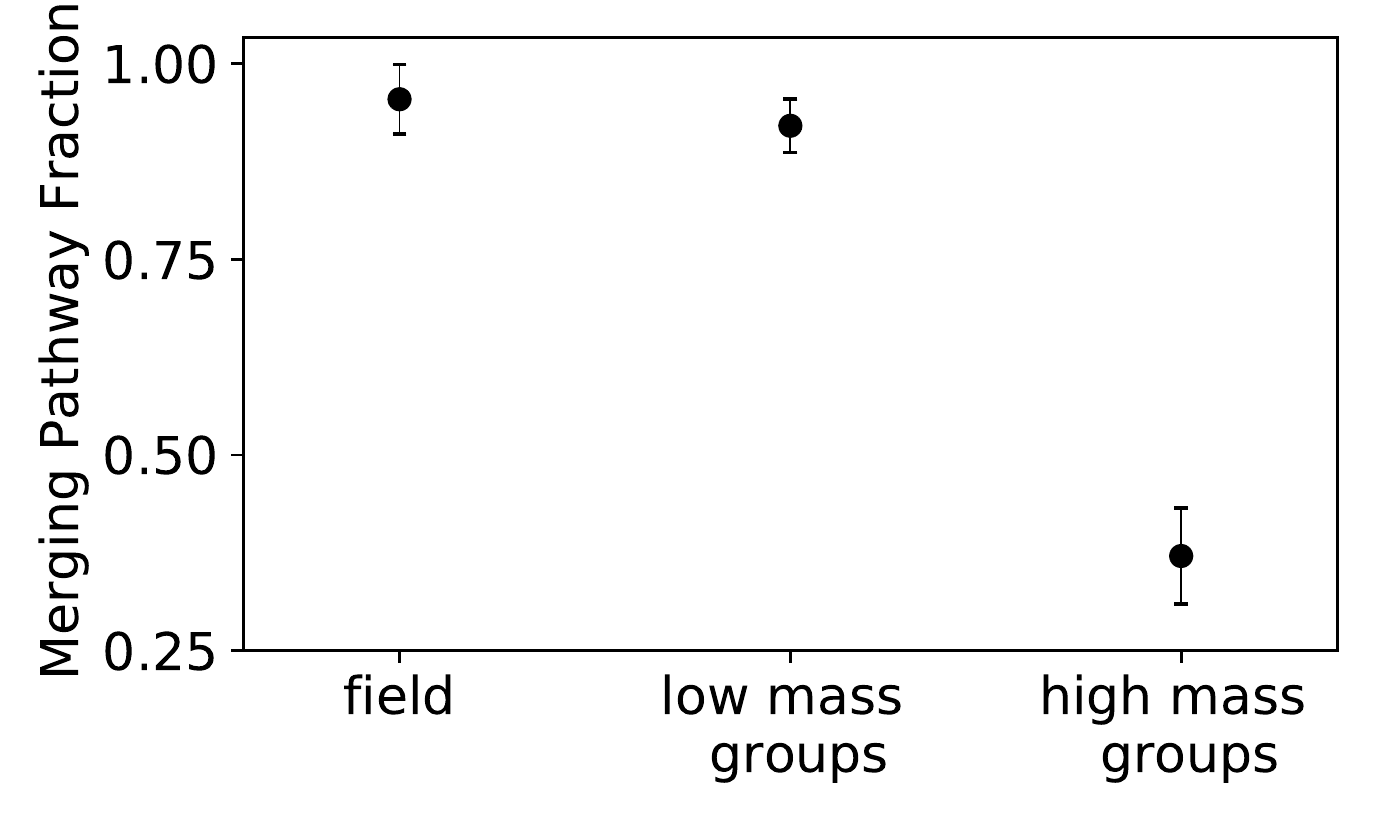}
   \caption{The relative occurrence of the merger pathway in different environments. S0s forming via the merger pathway dominate in low density environments while those forming via stripping dominate the high-density environments. Even though this is initially unsurprising since higher density environments are required for the infall pathway to occur, this interestingly suggests that there is very little pre-processing in these environments, i.e. very few galaxies within these groups entered the group after already transforming into an S0. However it should be noted that pre-processing may be more common in higher-density cluster environments.}
 \label{pathway_env}
 \end{figure}
 
 \subsection{Evolution of isolated S0s with no major events}
 
Six of the S0s in our sample featured no merger, interactions or infall events prior to their transformation. All of these galaxies were in very isolated environments when they transformed, with their halo mass making up between 92 and 95 percent of their host group. Five remain isolated up to the final snapshot while the remaining galaxy became associated with a larger group. Five of these galaxies follow an evolution very similar to the example shown in Figure~\ref{isolated}, where the gas and star forming activity gradually decreases and becomes more centrally concentrated. Given their isolation, this is likely the result of the lack of new gas coming in from the surrounding environment to replenish the galaxy as it forms new stars. The example here shows a strong central bar, however a bar isn't evident in all of these galaxies so a link to bar formation cannot be made. The remaining galaxy features a curve in its track through space and a sudden re-orientation of its spin axis; this suggests it may experience an interaction with another galaxy.

\subsection{Environmental Dependence}
\label{env}

Figure\ref{pathway_env} shows the fraction of S0s forming via the merger pathway as a function of environment. In the field and low density environments, the vast majority of S0s are forming via the merger pathway, while in the large groups the dominating pathway almost completely reverses to the infall pathway. This is initially not surprising, since a higher density environment is needed for ram pressure stripping to occur. However, it is interesting that very few S0s forming via the merger pathway have ended up in higher density environments. It has previously been proposed that many S0s in high density environments may be pre-processed \citep[e.g.][]{2015FrASS...2....4D}, i.e. they become S0s before entering the group environment, while our results here are indicating that most of these S0s actually enter in a star forming phase and transform during the infall itself. It should however be noted that due to the smaller size of IllustrisTNG-100, we are missing the highest density group environments, where hierarchical formation and pre-processing may be more prevalent. 

\subsection{Kinematics}

\begin{figure}
\includegraphics[width=0.9\columnwidth]{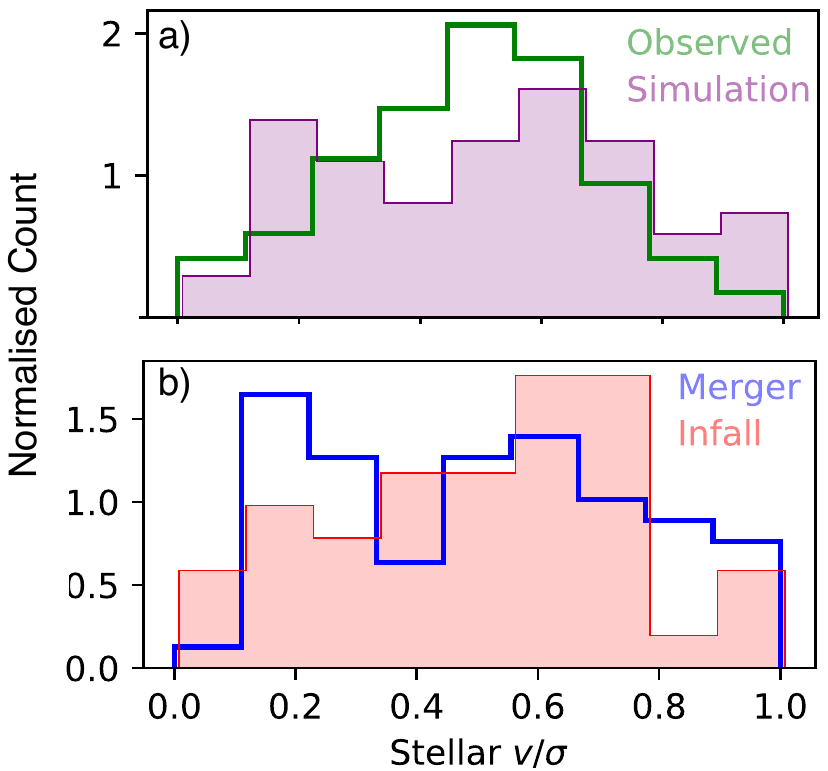}
   \caption{Distributions of the S0 stellar $v/\sigma$, using line-of-sight projections. a) compares the observed SAMI S0 sample (green) with the illustrisTNG sample (purple), showing that the range of rotational support in the simulation S0s is similar to observations. b) shows the $v/\sigma$ distribution for IllustrisTNG S0s in the merger-only group (blue) and the infall group (red). }
 \label{vsigma}
 \end{figure}

Figure~\ref{vsigma} a) compares the line-of-sight $v/\sigma$ distribution for S0s in illustrisTNG to the observed distribution in SAMI. The distribution in IllustrisTNG is less peaked, and in particular features a secondary peak at low $v/\sigma$. Otherwise, the range in the degree of rotational support is similar (with a two-sided KS test returning a p-value of 0.24). Figure~\ref{vsigma} b) compares the distribution of $v/\sigma$ for IllustrisTNG S0s in the merger-only group with those in the infall group. The infalling S0s are slightly skewed towards higher $v/\sigma$ values relative to the merger group, but this difference is not statistically signifiant (a two-sided KS test between the samples returns a p-value of 0.83). 

 In the two main groups, as well as those in the overlap region. S0s forming via the stripping pathway retain a higher degree of rotational support, as expected due to the less disruptive nature of this pathway. Those forming via mergers, meanwhile, have a higher degree of pressure support caused by the disruptive merger events. The range of $v/\sigma$ values for each pathway cover the regions of the rotationally supported and pressure-supported S0s in \citet{2020MNRAS.498.2372D}, supporting the claim made there that the large range in observed rotational support is a consequence of two dominant formation pathways contributing to the S0 population 

\subsection{The Elliptical - S0 Pathway}

\begin{figure*}
\includegraphics[width=1.4\columnwidth]{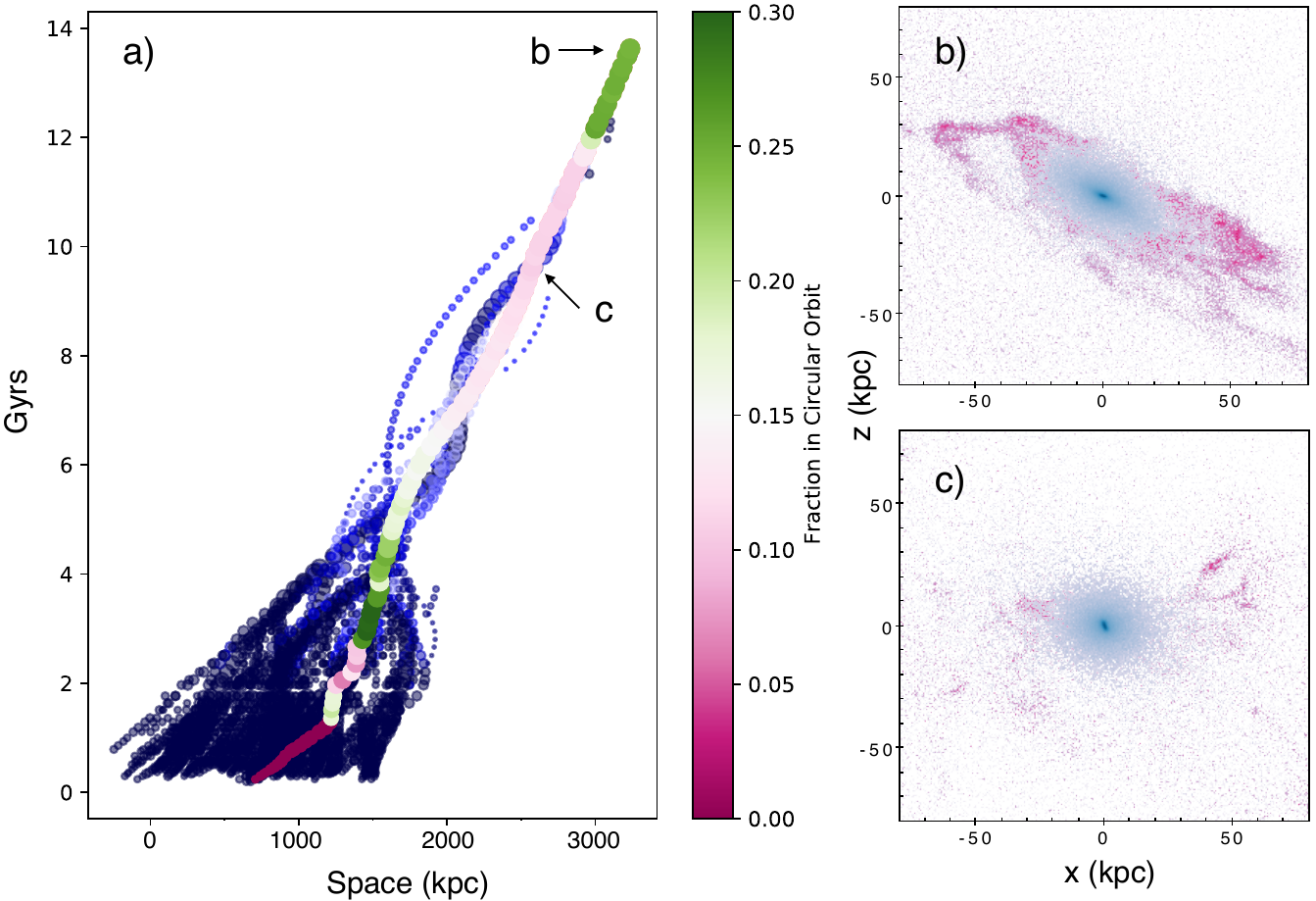}
   \caption{Example of an elliptical transforming into an S0 via a satellite merger. A) displays the merger tree of the galaxy, with the main track coloured by the circular-orbit fraction. After initially forming as a disk, the galaxy transitions to an elliptical and experiences an extended time in this state. When the merger occurs, the galaxy rapidly forms a disk component, indicated by the sudden jump in the circular orbit fraction. b) and c) show the stellar and gas morphology before and after the merger event, illustrating the embedded disk formed by the merger.}
 \label{cE}
 \end{figure*}

We searched our S0 sample for potential occurrences of an elliptical transforming into an S0. Such a pathway was recently put forward by \citet{2018MNRAS.477.2030D}, whereby a compact elliptical galaxy transformed into an S0 via a merger with a small gas-rich satellite. The influx of stars and gas from the satellite resulted in the formation of a disc structure within the original elliptical, transforming it into an S0 with an extended bulge. It was shown in Deeley et al (2020) that the properties of the final S0 were consistent with the group of S0s believed to arise from disruptive merger events. 

Using the colour index and circular-orbit fraction, we looked for galaxies which initially maintained an elliptical-like structure and then, following a merger event, experienced an increase in the circular orbit fraction indicating the formation of a disk. Of the three examples identified, we then looked at their stellar density distributions through each snapshot and found 1 case whereby the galaxy clearly followed the elliptical-merger-S0 pathway. The history of this galaxy, along with its morphology before and after the merger is shown in Figure~\ref{cE}. 

The fact that we only identify one example out of 127 S0s shows that this process is very rare in the Universe; from this one example we can only provide a rough upper limit of around 1 percent. Nevertheless, the fact that we were able to find such an example shows that this process can indeed occur in a cosmological context.

\subsection{Red Spirals}

\begin{figure*}
\includegraphics[width=1.7\columnwidth]{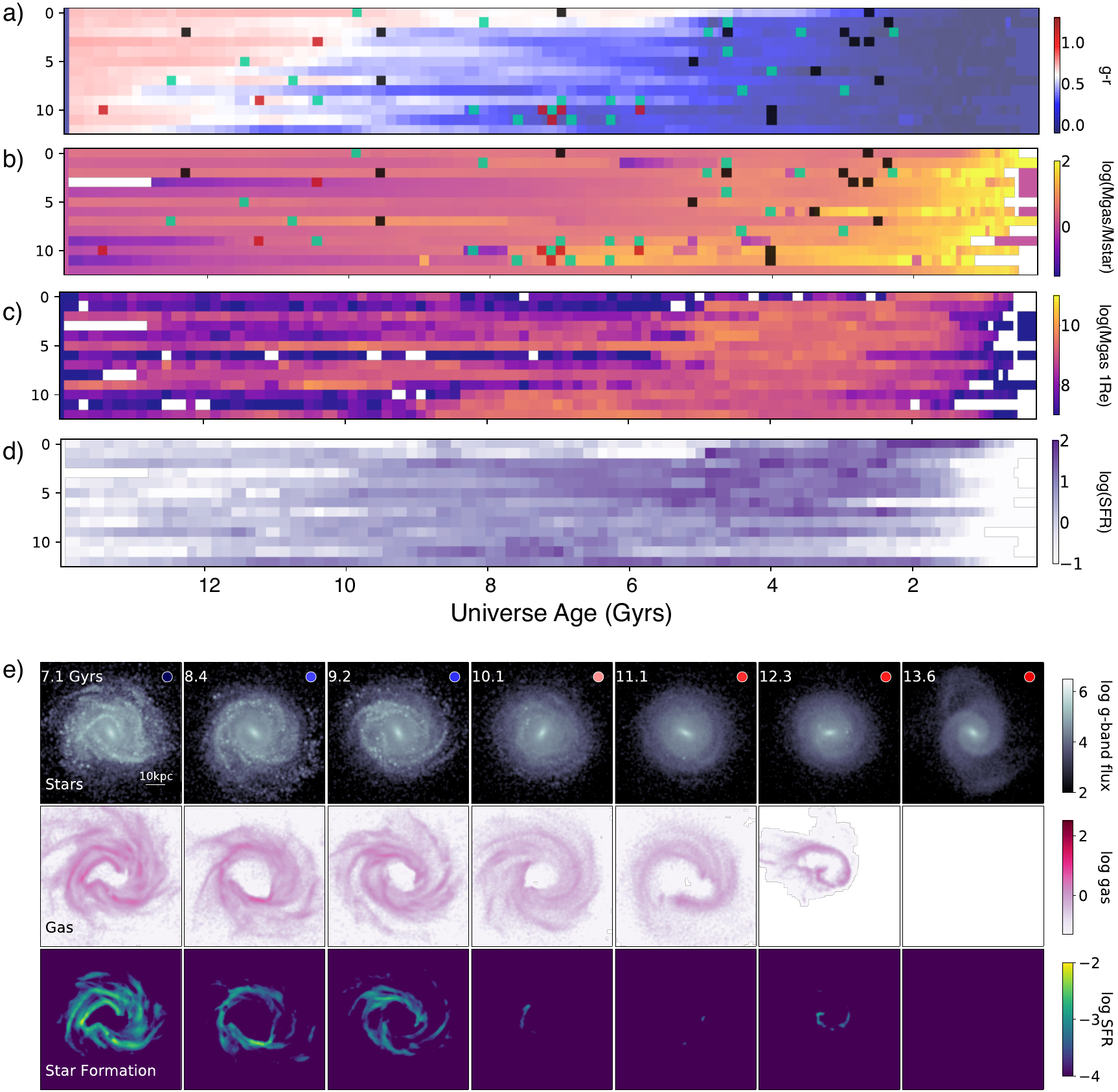}
   \caption{Formation histories of red spiral galaxies (early type spirals with a $g-r$ colour index redder than 0.7). a) shows the colour evolution, b) the total stellar-mass-weighted gas mass history, c) the gas mass within 1 Re, and d) shows the star formation history. A typical example of a red spiral is shown in e). The greatest contrast between these galaxies and the histories of S0s is the continued presence of star formation following the transition to a red galaxy, which may help to maintain the spiral structure for an extended time.  }
 \label{red_spirals}
\end{figure*}

An interesting subset of the spiral population consists of what are referred to as red spirals. These are galaxies which have prominent spiral arms yet feature very little star forming activity, giving them a red S0-like colour index. To see how such galaxies fit into the S0 picture being built here, we looked at the formation histories of early type spirals with a g-r colour index redder than 0.6, shown in Figure~\ref{red_spirals}. All of these galaxies appear to undergo a similar process to what is observed in the merger group; the central region becomes depleted of gas and the star formation activity is shifted out to the outer regions of the galaxy. However, the ring structure formed during this process is generally more diffused with arms/filaments leading off them, and once the main ring dissipates, star formation persists in isolated regions/arms of the galaxy for an extended period of time. This residual star formation may help to maintain the ring structure for an extended period of time, delaying the final transition to an S0. 

When looking at the star formation histories of S0s in Figure~\ref{histories} b), particularly in the merger group there are examples where star formation also continues for an extended time after the galaxy crosses the $g-r = 0.6$ line. Looking at the star distributions in some of these examples shows that they feature spiral arms during this phase, and would have therefore been considered as red spirals before losing this structure and becoming S0s. Therefore it appears that red spirals are galaxies going through a similar process as S0s in the merger group, with the less dominant ring phase and subsequent period of continued star forming activity resulting in an extended lifetime of the spiral structure.

\section{Discussion}

By following the formation histories of all S0s in our sample and looking at their events and gas evolution, we have found two very distinct groups forming via different pathways, along with a rare case of a third pathway. Here we further discuss each pathway and their relative abundance, building up a complete picture of S0 formation. 

\subsection{The support for two major pathways}

The two groupings of S0s identified here was based only on their final significant event before the transformation occurred. However, this separation is supported by additional differences between the two groups. Firstly, within the merger-only group, there are strong relations between the timing of the merger events and the time of transformation. This relationship is not as strongly evident within the infall group, where instead the timing of group infall events is correlated with the transformation time. This shows that the merger events do actually play a significant role in the transformation of the former while having a far less active role in the infall group, which are instead driven by the timing of the infall events. Secondly, S0s in the two groups have markedly different gas evolution; those in the merger group retain a significant amount of gas in their outer regions but become gas depleted in their central regions, while those in the infall group experience very rapid and complete gas stripping. Thirdly, the dynamics and activity of their spatial tracks have distinct appearances; those in the merger group have relatively constant tracks through space-time with significant merger activity, suggesting they are a central galaxy within their own small group halo, while those in the infall group have very dynamic tracks more indicative of a satellite galaxy. Thirdly, the merger group undergoes a phase where the gas becomes concentrated in a ring; while some S0s in the infall group also go through this phase along with pre-infall merger events, this is rapidly interrupted when the infall occurs.

\subsection{Stripped S0s}

We first focus on galaxies which transformed after a group infall event. We find that 37 percent of all S0s have undergone such an infall event.  

As we pointed out in Figure~\ref{histories}, the impact of the infall event is very dramatic. The transition from a blue star forming galaxy to a red passive galaxy occurs within 1-2 billion years of the initial infall, after it has passed near the centre of the cluster. The speed of this transition is driven by the very rapid and complete removal of gas from the galaxy and its surrounding halo once it enters the cluster's R200 radius, quickly brining star formation to a halt. During this stage, gas trails behind the galaxy giving it the appearance of a jellyfish. Due to the outside-in nature of this gas stripping, star formation continuous in the centre during the initial stages of the infall, producing a younger bluer core relative to the disk. Indeed, S0s in clusters have recently been observed to feature younger cores and older disks \citep{2012MNRAS.422.2590J}.

Evidently, all of these galaxies are located in high-density environments in the present day. What is notable however, is that nearly all of the S0s in these environments underwent an infall event while they were still in a star-forming phase, i.e. very few of them were originally located within these environments. This indicates that most S0s in these environments were transformed by the infall into that environment rather than pre-processed. It should be quickly noted however that IllustrisTNG-100 lacks truly massive galaxy clusters, where pre-processing may be a larger contributor to the group S0 population due to their more extensive hierarchical formation. This could be investigated using the lower resolution, larger volume IllustrisTNG-300 simulation which contains many halos with high group masses.

\subsection{Merger-only S0s}

For S0 galaxies which only experienced merger events in their past, the relationship between the timing of the merger event and the time of transformation suggests that these mergers may be playing an active role in the transformation. 

The first mergers occur early in the galaxy's formation, during their initial buildup phase. These mergers tend to lower the gas/star ratio, and are therefore likely accelerating the exhaustion of gas and hence reducing the period of star formation. The second period of mergers occurs an average of 4 billion years before the transformation, and is associated with a significant drop in gas within 1 half-mass radii soon after. Star formation activity continues for a period after the merger, fuelled by gas from the merging galaxy and falling in from the surrounding halo, until these supplies are exhausted. How these merger events differ from those which do not trigger a transformation to an S0 will be investigated in future work.

Surprisingly, there is in most cases a significant time lag between the merger events and the time of transformation, often exceeding 4 Gyrs. Visual investigations of the morphologies of the galaxies, using the g-band stellar luminosity, show that after the merger event strong spiral arms often reform, before eventually weakening and disappearing to leave a smooth, featureless disk. Separating the stellar population into those which formed before and after the merger event reveals that the pre-merger population is significantly disrupted by the merger, erasing the previous spiral structure and resulting in a thick, smooth disk. Stars formed after the merger are highly concentrated near the central plane in a thin disk with strong spiral arms. We postulate that the re-formed spiral structure persists while the remaining gas is able to fuel star formation, after which the interaction between the thick disk of pre-merger stars and the thin disk of newly formed stars disrupts the spiral structure and shortens its lifetime.

\begin{figure*}
\includegraphics[width=1.6\columnwidth]{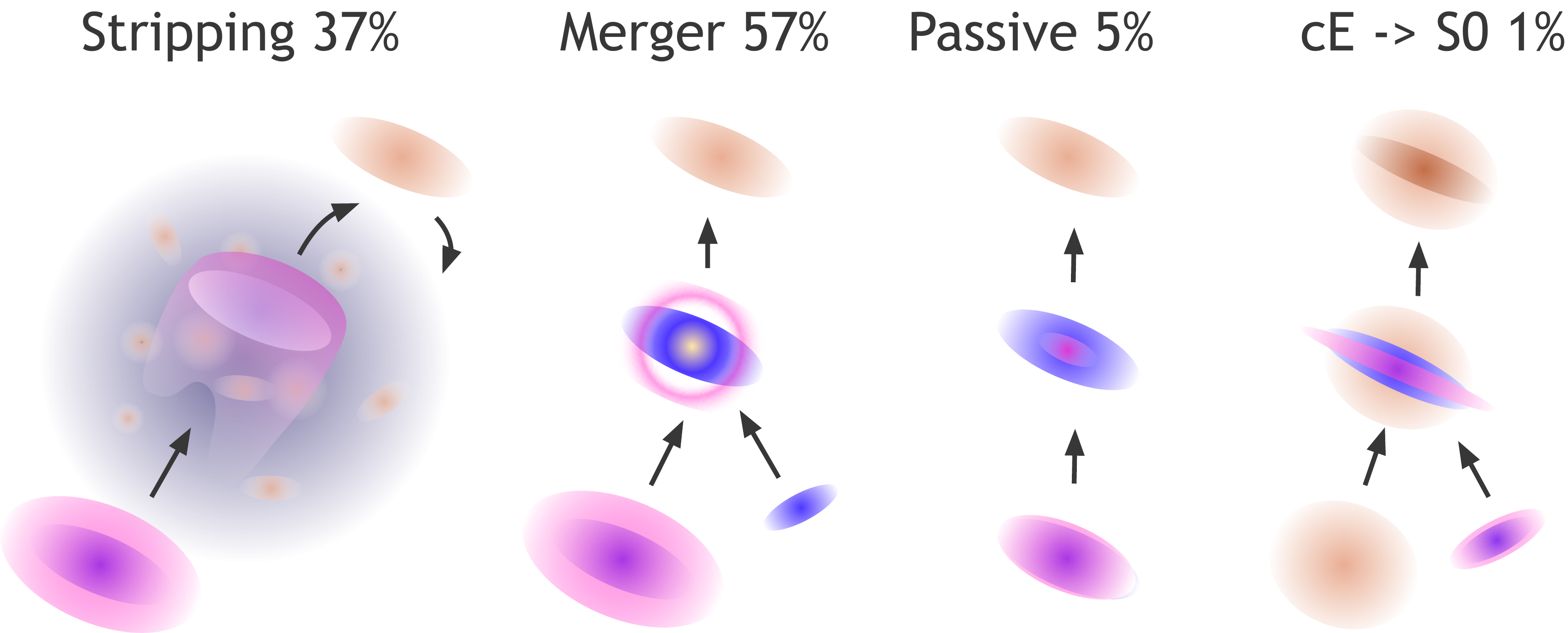}
   \caption{The overall picture of S0 formation presented in this work. Blue - young stellar populations, orange - older stellar populations, pink - gas, purple - intra-cluster medium.}
 \label{picture}
\end{figure*}

This time lag is larger than typically seen in other simulations. \citet{2015A&A...573A..78Q} found that merger remnants resembled an S0 galaxy within less than 3 Gyrs, however at this stage the galaxies were still bluer than observed due to residual star formation, and another 1-2 Gyrs would be needed for them to redden to observed S0 colours. In addition, this residual star formation was located in the core leaving passive disks, while in many of our merger galaxies it is located within the disk, giving rise to the remnant spiral structure. Therefore, the remanent spiral arms combined with the delay in colour change may explain the longer transformation time seen here. Isolated simulations also don't typically include the extended gas halo (such as the gas halo shown in Figure~\ref{isolated}); infalling gas from the extended halo may also act to extend the residual star formation and the persistence of spiral arms, since the star formation rate we see in the disk would otherwise rapidly deplete the gas contained within the disk alone.

In the vast majority of these galaxies, a ring of star formation forms after the merger event, when the transformation in colour is occurring. This often (but not always) corresponds to the formation of a central bar. The formation of this ring removes gas from the central regions of the galaxy and concentrates star formation activity in a very narrow region further out. This brings and end to star formation within the central region, beginning the transformation of the galaxy from blue to red. Being concentrated into the ring, gas is consumed by star formation and the ring breaks up, eventually leaving no star formation within the entire disk. This process would result in an older core and younger disk, as is observed in S0s within lower density environments \citep{2017MNRAS.466.2024T}.

As noted above, $\rm H\alpha$ rings have been observed in a small fraction of S0s and early type galaxies within SAMI and CALIFA. The occurrence rate of the rings seen here, within a single late snapshot of the simulation closely matches these observations. It therefore appears that these S0s are galaxies which are actively undergoing (or recently completed) the transformation from a star forming spiral. The low occurrence rate of observed rings, despite appearing at some stage in most of these S0s, can be explained by the short lifetime of these structures (typically around 1 Gyr). Targeting such galaxies will therefore allow us to more closely observe this process as it occurs. Such rings also occur in the histories of some S0s in the infall group, where the galaxies have experienced a merger before the infall event stripped out the remaining gas. 

\subsection{Gas exhaustion}

The six galaxies which cannot be placed into either of the two main groups above appear to transform into S0s purely due to the exhaustion of gas. All of these galaxies are very isolated when the transformation occurs, being the only significant object within their host group. It is therefore likely that these galaxies have very little gas in their surroundings, meaning that the gas cannot be replenished as it is used up in star formation and they eventually run out of star forming material. 

\subsection{Overall picture}

From the results presented here, we can then construct an overall picture of the formation of S0 galaxies and compare it to observations (illustrated in Figure~\ref{picture}). We see that S0 formation mainly occurs by either a stripping event caused by a group infall or a merger event. These two pathways have nearly equal importance, with the first leading to around 37 percent of S0s and the second to 57 percent. Many infalling galaxies also clearly experience earlier merger events, and indeed appear to have been transforming into S0s prior to infall. However the complete gas stripping of the infall event is significantly more rapid and dramatic, and therefore has a far more significant impact on the subsequent evolution of the galaxy. A small subset of very isolated S0s do not experience any significant events, and transform into S0s through exhaustion of their limited gas supply.

 In addition, there are rare occurrences of an elliptical galaxy transforming into an S0 via a gas-rich satellite merger, making up a small but non-zero fraction of the population. Due to finding only one case within our sample, we can only provide a rough upper limit to this pathways' contribution to the S0 population of around 1 percent. Recent observational work by \citep{2015ApJ...804...32G} looking at the bulges of disk galaxies also suggests that this pathway is contributing to the final population, albeit at a potentially higher occurrence rate than found here. 
 
Looking at the average rotational support of these two main groups showed that the infalling S0s retain a larger degree of rotational support on average compared to those which have formed via a merger. This is due to the highly disruptive nature of these mergers, which heat up the disk and impart a higher degree of random motion. Indeed, when we instead separate the S0 population by wether or not they have undergone a major merger, the distinction in rotational support becomes greater. The reason why we see a higher average degree of rotational support in the infall group is due to the examples of infalling galaxies which are able to transform into S0s without a merger disruptive merger event, thereby retaining a higher degree of rotational support. 

Since these infalling S0s are located more frequently in lower mass group environments, we therefore also expect to see an environmental dependence of rotational support with higher mass environments featuring higher degrees of rotational support. Indeed, this is what has been observed within the GAMA region of the SAMI S0 sample \citep{2020MNRAS.498.2372D}. However, the comparison between our full simulation sample and the observed S0 sample show that while they are marginally consistent, there may be discrepancies in the kinematics derived from the simulation which will need to be considered in future comparisons.

Due to the in-falling S0s also experiencing merger activity, and the variation in merger ratios of the other population, we end up with a wide range of rotational support for each pathway. This means that it is very difficult to cleanly separate them into two separate populations within observations by their kinematics alone. 

As was seen from the morphological classification, there appears to be a lack of galaxies with very low bulge-to-disk ratios. These discrepancies in bulge sizes may contribute to the kinematic discrepancies noted above, since larger bulges would reduce the $v/\sigma$ ratio. While this also impacted the separation of early and late-type spiral galaxies, it is not expected to significantly impact the visual classification of S0 galaxies. Given that the bulge can influence the quenching of star formation \citep{2013ApJ...776...63F}, this discrepancy may lead to additional low-mass spirals transforming into S0s in the simulation, most likely through the merger or passive pathways given the more gradual nature of these processes. However, we note that the agreement between the S0 fraction in the observation and simulation samples (Figure~\ref{group_frac}) suggests that the contribution of additional S0s from this affect may be minimal.

\section{Summary and Conclusions}

In order to further investigate the formation pathways of S0 galaxies, we identified S0s within the IllustrisTNG simulation and traced their histories back through time. Based on the events each galaxy experienced prior to its transformation, we identified two main groups; the first group (37 percent of all S0s) experienced a significant infall event prior to transformation while the second group (57 percent) experienced a significant merger event. The infall events were found to trigger a very rapid transformation, with the galaxy turning into an S0 within 1-2 billion years of the event, while we found the merger events to transform the galaxy over a longer period of time. In addition, 6 percent of S0s formed via the passive fading of spiral galaxies in very isolated environments with no merger or infall events in their past, likely due to the limited amount of fresh star-forming material falling in from the surrounding environment. Discrepancies in the bulge-to-total ratio within the simulation may influence the final relative contributions of these pathways, however we don't expect this to significantly impact our main conclusions.

Looking at the gas and star formation histories of these S0s, those forming via group infall events experience very rapid and complete gas stripping from the outside in, which is responsible for a rapid halt of star formation activity in these galaxies. The merger group, meanwhile, experience a significant depletion of their internal gas within their central regions after the merger. After the merger, spiral arms reform and can continue to persist for over 4 Gyrs while star formation activity continues. The remaining gas becomes concentrated in a ring structure with a typical lifetime of around 1 Gyr, and is eventually consumed by star formation. Once the gas is exhausted, the spiral arms dissipate and the galaxy takes on the morphology of an S0.  

We also find one example of the pathway proposed in \citet{2018MNRAS.477.2030D}, where an elliptical galaxy experiences a merger with a small gas-rich satellite, forming a disk and transforming into an S0. This demonstrates that this scenario can occur naturally in a cosmological context. However, finding only a single example shows this is a very rare occurrence, and allows us only to place a rough upper limit for their contribution to the S0 population of around 1 percent. 

Finally, we determined the degree of rotational support of S0s in each group, finding that the infall group retain a larger degree of rotational support relative to the merger group. While noting that differences in the distribution of $v/\sigma$ between the simulation and observations indicates a potential discrepancy, the range in $v/\sigma$ closely corresponds to that observed in the SAMI survey, supporting the claim that the rotationally-supported S0s were formed via faded spiral pathways and the more pressure-supported S0s were formed via merger events.  However, due to in falling S0s also experiencing varying amounts of merger activity as well as the variations in the mergers themselves, there is significant overlap in the final $v/\sigma$ measurements making the two groups difficult to separate in observations.

\section*{Acknowledgements}

The authors thank Joel Pfeffer for helpful comments on the manuscript. This work was supported through the Australian Research Council's Discovery Projects funding scheme (DP170102344). Support for AMM is provided by NASA through Hubble Fellowship grant \#HST-HF2-51377 awarded by the Space Telescope Science Institute, which is operated by the Association of Universities for Research in Astronomy, Inc., for NASA, under contract NAS5-26555. 

\section*{Data Availability} 
The data underlying this article were accessed from the publicly available IllustrisTNG-100 simulation \citep{2018MNRAS.481.2299S}, available at https://www.tng-project.org/data. The derived data generated in this research will be shared on reasonable request to the corresponding author.

%%%%%%%%%%%%%%%%%%%%%%%%%%%%%%%%%%%%%%%%%%%%%%%%%%


\bibliographystyle{mnras}

%%%%%%%%%%%%%%%%%%%%%%%%%%%%%%%%%%%%%%%%%%%%%%%%%%

%%%%%%%%%%%%%%%%% APPENDICES %%%%%%%%%%%%%%%%%%%%%

\appendix

%%%%%%%%%%%%%%%%%%%%%%%%%%%%%%%%%%%%%%%%%%%%%%%%%%

% Don't change these lines
\bsp	% typesetting comment
\label{lastpage}
\end{document}